\documentclass[12pt]{article}
\pdfoutput=1

\usepackage{amsmath,amssymb,amsfonts,graphicx,amsfonts}
\usepackage{textcomp,mathcomp}
\usepackage{color,xcolor}
\usepackage[hidelinks]{hyperref}
\usepackage{ascmac}
\usepackage{physics}

\allowdisplaybreaks[4]

\date{}
\begin{document}
\begin{flushright}
\today\\
%June 19, 2024
\end{flushright}

\vspace{0.1cm}

\begin{center}

  {\Large Operator algebra, quantum entanglement, and emergent geometry} \\
  \vspace{2mm}
{\Large   from matrix degrees of freedom}\\

\end{center}
\vspace{0.1cm}
\vspace{0.1cm}
\begin{center}

Vaibhav Gautam$^{a}$, Masanori Hanada$^{a}$, and Antal Jevicki$^b$

\end{center}
\vspace{0.3cm}

\begin{center}

{\small

$^a$ School of Mathematical Sciences, Queen Mary University of London\\
Mile End Road, London, E1 4NS, United Kingdom\\
\vspace{1mm}
$^b$  Brown Theoretical Physics Center, Brown University\\
340 Brook Street, Providence, RI 02912, United States\\

}

\end{center}

\vspace{1.5cm}

\begin{center}
  {\bf Abstract}
\end{center}

For matrix model and QFT, we discuss how dual gravitational geometry emerges from matrix degrees of freedom (specifically, adjoint scalars in super Yang-Mills theory) and how operator algebra that describes an arbitrary region of the bulk geometry can be constructed. We pay attention to the subtle difference between the notions of wave packets that describe low-energy excitations: \textit{QFT wave packet} associated with the spatial dimensions of QFT, \textit{matrix wave packet} associated with the emergent dimensions from matrix degrees of freedom, and \textit{bulk wave packet} which is a combination of QFT and matrix wave packets. In QFT, there is an intriguing interplay between QFT wave packet and matrix wave packet that connects quantum entanglement and emergent geometry. We propose that the bulk wave packet is the physical object in QFT that describes the emergent geometry from entanglement. This proposal sets a unified view on two seemingly different mechanisms of holographic emergent geometry: one based on matrix eigenvalues and the other based on quantum entanglement. Further intuition comes from the similarity to a traversable wormhole discussed as the dual description of the coupled SYK model by Maldacena and Qi: the bulk can be seen as an eternal traversable wormhole connecting boundary regions. 

\newpage
\tableofcontents

%%%%%%%%%%%%
%%%%%%%%%%%%
\section{Introduction}
\hspace{0.51cm}
%%%%%%%%%%%%
%%%%%%%%%%%%
In this paper, we discuss how holographic emergent geometry can be described by matrix degrees of freedom of the dual QFT or matrix model, polishing basic ideas provided in Refs.~\cite{Hanada:2021ipb,Hanada:2021swb,Gautam:2022akq}.

Matrix degrees of freedom in gauge theory play important roles in holographic emergent geometry~
\cite{Hanada:2016pwv,Hanada:2018zxn,Hanada:2021ipb,Hanada:2021swb,Gautam:2022akq}. For example, a small black hole in AdS can be described by a partially-deconfined state in gauge theory~\cite{Hanada:2016pwv,Hanada:2018zxn}, and entanglement between the confined sector and deconfined sector explains the Bekenstein-Hawking entropy as von Neumann entropy~\cite{Gautam:2022akq}. The use of wave packet in the space of matrix degrees of freedom~\cite{Hanada:2021ipb,Hanada:2021swb} is crucial in resolving a puzzling conceptual issue raised by Polchinski~\cite{Polchinski:1999br} and Susskind~\cite{Susskind:1998vk}.
To read the emergent geometry from matrices, one has to carefully distinguish two different notions of ``target space". In the case of the D0-brane matrix model, there is a space $\mathbb{R}^9$ in which string theory lives, and a space of matrix coordinates $\mathbb{R}^{9N^2}$ (i.e., there are nine matrices and $N^2$ real numbers are needed to specify each of them). Wave packets live in $\mathbb{R}^{9N^2}$, and from wave packets, we can extract the information of $\mathbb{R}^9$. The confusion in the past came from the ignorance of this point.\footnote{It is also possible to make a wave packet in $\mathbb{R}^9$ as a superposition of wave packets in  $\mathbb{R}^{9N^2}$.} 

Recently, holographic emergent geometry was discussed in terms of von Neumann algebra associated with effective field theory around semiclassical geometry on the gravity side~\cite{Witten:2018zxz,Leutheusser:2021qhd,Leutheusser:2022bgi}. To make further progress, it is important to understand how this approach and matrix degrees of freedom are related. In this paper, we tackle this problem. We will study the matrix model first, then generalize the idea to QFT. 

As a byproduct, we find a simple connection between the two seemingly different mechanisms of emergent geometry: one from matrix degrees of freedom~\cite{Banks:1996vh,Hanada:2016pwv,Hanada:2021ipb,Hanada:2021swb,Gautam:2022akq} and the other from quantum entanglement~\cite{Maldacena:2001kr,VanRaamsdonk:2010pw}. This is achieved by paying attention to subtle differences in the notions of wave packets that describe low-energy excitations: \textit{QFT wave packet} associated with the spatial dimensions of QFT, \textit{matrix wave packet} associated with the emergent dimensions from matrix degrees of freedom, and \textit{bulk wave packet} which is a combination of QFT and matrix wave packets. 
The duality dictionary we propose for AdS$_5$/CFT$_4$ correspondence is the following:\\
\begin{screen}
\begin{center}
\begin{tabular}{ccc}
QFT & & gravity\\
\vspace{-4mm}\\
\hline
\vspace{-3mm}\\
bulk wave packet & $\longleftrightarrow$ & closed string\\
\vspace{-5mm}\\
$\Big\downarrow$\scriptsize{condense} & \scriptsize{dual} & $\Big\downarrow$\scriptsize{condense}\\
\vspace{-4mm}\\
global matrix wave packet & $\longleftrightarrow$ & D3-brane\\
\end{tabular}
\end{center}
\end{screen}
\vspace{1mm}\\
The definition of bulk wave packet and global matrix wave packet will be given in Sec.~\ref{sec:QFT}. 
To create a bulk wave packet sitting close to the center of the spatial dimensions that emerge from matrix degrees of freedom, it is necessary to entangle a large region on the boundary. Thus, we identify a physical object in QFT that describes the emergent geometry from entanglement: a bulk wave packet is associated with quantum entanglement on the boundary. 

This paper is organized as follows.
In Sec.~\ref{sec:matrix-geometry-review}, we review wave packets in the matrix model. We will see how emergent geometry can be encoded into matrices. In Sec.~\ref{sec:matrix_model}, the algebra of small excitations in the matrix model is discussed. Generalization of these ideas to QFT is discussed in Sec.~\ref{sec:QFT}. We mainly consider AdS$_5$/CFT$_4$ correspondence. In Sec.~\ref{sec:D3-brane}, the global matrix wave packet and D3-brane are discussed. QFT wave packet and bulk wave packet are introduced in Sec.~\ref{sec:QFT-and-bulk-wave-packets}, and their properties are discussed in Sec.~\ref{sec:geometry-bulk-wave-packet} in the context of holographic duality. Sec.~\ref{sec:conclusion} is devoted to conclusions and discussions. 
%%%%%%%%%%%%
%%%%%%%%%%%%
\section{Review of wave packet in matrix model}\label{sec:matrix-geometry-review}
\hspace{0.51cm}
%%%%%%%%%%%%
%%%%%%%%%%%%
In this section, we review the emergent geometry in the matrix model following Refs.~\cite{Hanada:2021ipb,Hanada:2021swb,Gautam:2022akq}. 
A more detailed presentation can be found in Sec.~2 of Ref.~\cite{Hanada:2023rlk}.

There was a good old idea of emergent geometry based on the identification of eigenvalues of matrices and locations of emergent space, for low-energy dynamics of D-branes~\cite{Witten:1995im}, supermembrane~\cite{deWit:1988wri}, and the Matrix Model of M-theory~\cite{Banks:1996vh}. However, a puzzling conceptual issue~\cite{Polchinski:1999br,Susskind:1998vk} regarding the application of this idea to gauge/gravity duality a la Maldacena~\cite{Maldacena:1997re} distracted researchers from pursuing this direction. (We explain this issue briefly in Appendix~\ref{sec:Polchinski-puzzle}.) This puzzle was resolved by noticing the importance of the wave packet in the space of matrix degrees of freedom\cite{Hanada:2021ipb,Hanada:2021swb}, which we call \textit{matrix wave packet} in this paper. 

In this section, we will consider a bosonic SU($N$) nine-matrix model analogous to the BFSS matrix model. Each matrix has $N^2$ bosonic degrees of freedom, and hence, the wave functions live on $\mathbb{R}^{9N^2}$. The matrix wave packet is defined on $\mathbb{R}^{9N^2}$. From the matrix wave packet, the information of nine-dimensional space on the gravity side can be derived. 

Based on the materials reviewed in this section, we will discuss the description of an arbitrary region of emergent geometry from the matrix model in Sec.~\ref{sec:matrix_model}. In Sec.~\ref{sec:QFT}, we discuss the generalization to QFT. There, it is important to understand the relationship between two notions: \textit{matrix wave packet}  and \textit{QFT wave packet}. By combining matrix wave packet and QFT wave packet, we will reach the notion of \textit{bulk wave packet} that plays the fundamental role in AdS$_5$/CFT$_4$ correspondence.
%%%%%%%%%%%%
%%%%%%%%%%%%
\subsection{Gauge-invariant Hilbert space and extended Hilbert space}\label{sec:H_inv-vs-H_ext}
\hspace{0.51cm}
%%%%%%%%%%%%
%%%%%%%%%%%%
The matrix wave packet can be defined regardless of the details of the models, although dual interpretation in weakly-coupled gravitational theory can exist only for special kinds of models. Therefore, we consider the Hamiltonian interpolating between the Gaussian matrix model and the bosonic part of the BFSS matrix model as an example:\footnote{
This normalization corresponds to the Lagrangian $L={\rm Tr}\left(\frac{m}{2}(D_tX_I)^2-\frac{m\omega^2}{2}X_I^2+\frac{g^2}{4}[X_I,X_J]^2\right)$. Another common normalization is $L=N{\rm Tr}\left(\frac{m}{2}(D_tX'_I)^2-\frac{m\omega^2}{2}X_I^{\prime 2}+\frac{\lambda}{4}[X'_I,X'_J]^2\right)$, where $X'_I=N^{-1/2}X_I$. The 't Hooft scaling is $\langle\mathrm{Tr}(X'_{I_1}X'_{I_2}\cdots X'_{I_k})\rangle\sim N$ and $\langle\mathrm{Tr}(X_{I_1}X_{I_2}\cdots X_{I_k})\rangle\sim N^{1+k/2}$. 
}
\begin{align}
\hat{H}
=
{\rm Tr}\left(
\frac{1}{2m}\hat{P}_I^2
+
\frac{m\omega^2}{2}\hat{X}_I^2
-
\frac{g^2}{4}[\hat{X}_I,\hat{X}_J]^2
\right)\, . 
\label{eq:toy-Hamiltonian}
\end{align}
Here, $\hat{X}_{I}$ and $\hat{P}_J$ are operators describing Hermitian matrices, and hence, $(\hat{X}_{ij})^\dagger=\hat{X}_{I,ji}$ and $(\hat{P}_{ij})^\dagger=\hat{P}_{I,ji}$, where $i,j=1,2,\cdots,N$ and $^\dagger$ denotes Hermitian conjugation as operator acting on the Hilbert space. Trace is over matrix indices (as $N\times N$ matrices), and  $I,J$ run from 1 to 9. 
The Gaussian matrix model is obtained at $g^2=0$, and in the strong 't Hooft coupling limit $\lambda=g^2N\to\infty$ or $\omega^2\to 0$ the bosonic part of the BFSS matrix model is obtained. 
The canonical commutation relation is
\begin{align}
[\hat{X}_{I,ij},\hat{P}_{J,kl}]
=
i\delta_{IJ}\delta_{il}\delta_{jk}\, . 
\end{align}

To describe the matrix model through the quantum states, it is crucial to discern between the gauge-invariant Hilbert space ${\cal H}_{\rm inv}$, comprising solely of gauge-singlet states, and the extended Hilbert space ${\cal H}_{\rm ext}$ that encompasses non-singlet states as well.
In ${\cal H}_{\rm ext}$, we can employ the coordinate eigenstates $|X\rangle$ satisfying $\hat{X}_{I,ij}|X\rangle=X_{I,ij}|X\rangle$ for all $9N^2$ combinations of $(I,i,j)$ as the basis. Alternatively, we can utilize the momentum eigenstates $|P\rangle$ as the basis.
It is noteworthy that operators $\hat{X}_{I,ij}$ and $\hat{P}_{I,ij}$ are defined within this extended Hilbert space because they transform as adjoint representation under the SU($N$) transformation.
In the following, we will employ wave-packet states as an over-complete basis for ${\cal H}_{\rm ext}$.

The utilization of the extended Hilbert space finds natural justification when employing the projector from $\mathcal{H}_{\rm ext}$ to $\mathcal{H}_{\rm inv}$, which corresponds to integrating over the Polyakov loop~\cite{Hanada:2020uvt,Hanada:2021ipb}. The partition function at finite temperature can be expressed in two different yet equivalent ways. The first expression is
\begin{align}
Z(T)
&=
{\rm Tr}_{{\cal H}_{\rm inv}}\left(
e^{-\hat{H}/T}
\right). 
\label{eq:Z-H-inv-MM}
\end{align}
Because of this expression, it is common to hear the phrase ``physical states are gauge-singlets." The subsequent expression delves into the extended Hilbert space that encompasses non-singlet modes:
\begin{align}
Z(T)
&=
\frac{1}{{\rm vol}G}\int_Gdg
{\rm Tr}_{{\cal H}_{\rm ext}}\left(
\hat{g}
e^{-\hat{H}/T}
\right)
\equiv
{\rm Tr}_{{\cal H}_{\rm ext}}\left(
\hat{{\cal P}}
e^{-\hat{H}/T}
\right)\, . 
\label{eq:Z-H-ext-MM}
\end{align}
Here, $G={\rm SU}(N)$ represents the gauge group, and $g$ and $\hat{g}$ are a group element and its representation acting on the extended Hilbert space ${\cal H}_{\rm ext}$. The Haar measure is utilized for the integral.
The second expression is directly linked to the path integral involving the temporal gauge field $A_t$, where $g\in G$ corresponds to the Polyakov loop. In this second expression, it can be stated that ``states connected by a gauge transformation are identified," meaning that $|\phi\rangle$ and $\hat{g}|\phi\rangle$ are indistinguishable, analogous to what is observed in the classical theory or in the path-integral formulation.
$
\hat{{\cal P}}
\equiv
\frac{1}{{\rm vol}G}\int_G dg
\hat{g}
$
is a projection operator from ${\cal H}_{\rm ext}$ to ${\cal H}_{\rm inv}$ that satisfies $\hat{{\cal P}}^2=\hat{{\cal P}}$. 
From a state $|\phi\rangle\in{\cal H}_{\rm ext}$, which can be non-singlet, we can obtain a singlet $\hat{{\cal P}}|\phi\rangle\in{\cal H}_{\rm inv}$. 
%%%%%%%%%%%%%%%
%%%%%%%%%%%%%%%
\subsection{Matrix wave packet in $\mathbb{R}^{9N^2}$}\label{sec:wave-packet}
\begin{figure}[htbp]
  \begin{center}
   \includegraphics[width=80mm]{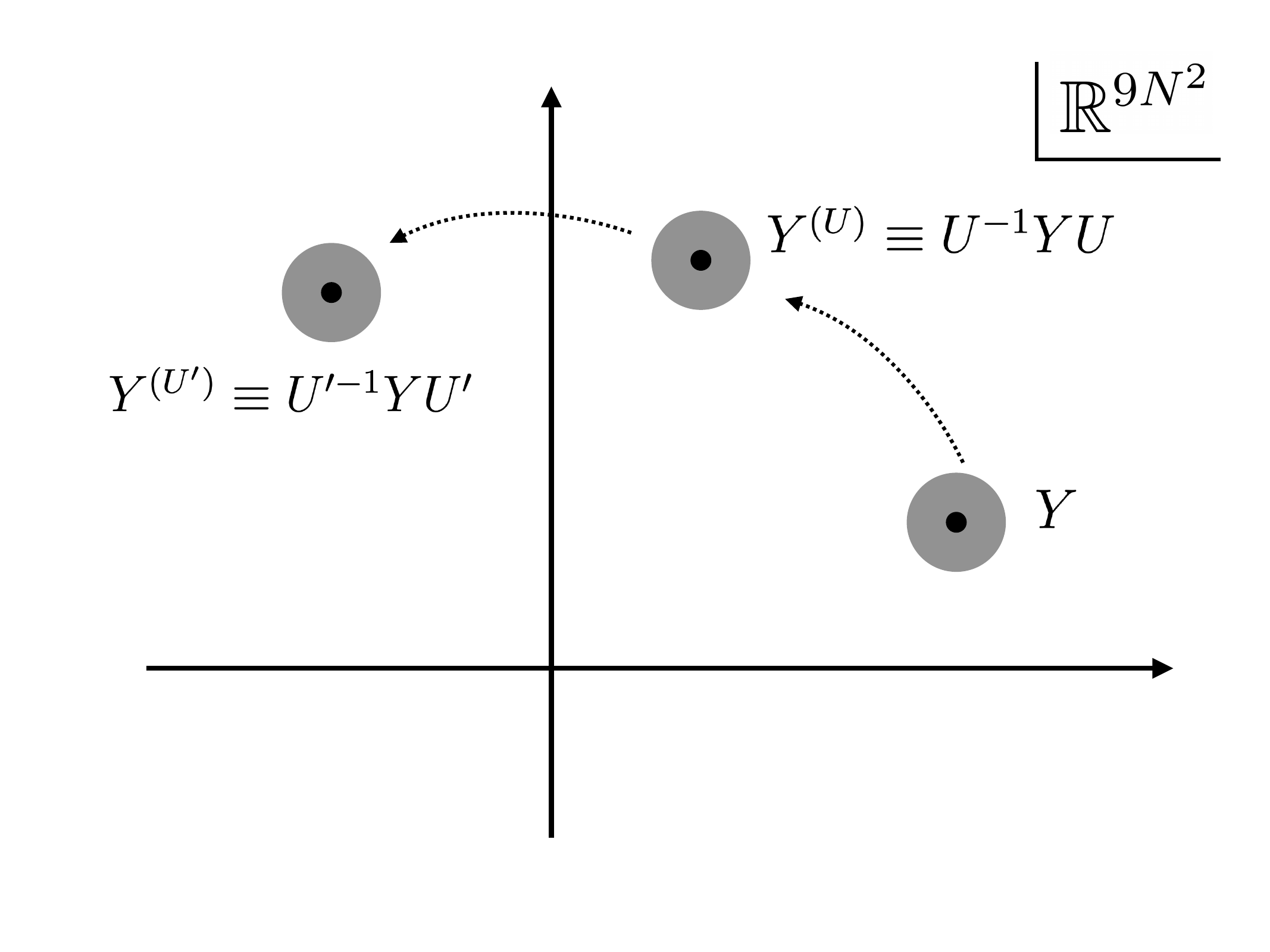}
  \end{center}
  \caption{
Wave packets in coordinate space $\mathbb{R}^{9N^2}$ are depicted. The black points represent the centers of wave packets ($Y$, $Y^{(U)}$, and $Y^{(U')}$), while the gray disks represent the wave packets, specifically the regions where the wave functions are not vanishingly small.
Under gauge transformation, the center of the wave packet shifts as $Y\to Y^{(U)}=U^{-1}YU$, while the shape of the wave packet remains unchanged. The center of the wave packet characterizes the geometry comprising D-branes and strings. This figure is adapted from Ref.~\cite{Gautam:2022akq}.
  }\label{fig:wave-packet}
\end{figure}

%%%%%%%%%%%%%%%
%%%%%%%%%%%%%%%
Low-energy states can be written as superpositions of \textit{wave packets in the space of matrices} that is $\mathbb{R}^{9N^2}$. 
We use the name \textit{matrix wave packet} to emphasize that this is a wave packet in $\mathbb{R}^{9N^2}$ and not in $\mathbb{R}^{9}$. 
We consider a wave packet around $Y_I$ in coordinate space and $Q_I$ in momentum space.
Such a wave packet naturally characterizes the geometry formed by D-branes and open strings~\cite{Hanada:2021ipb,Hanada:2021swb}, following the approach introduced in Ref.~\cite{Witten:1995im}.
A straightforward method to obtain a low-energy wave packet is to identify a state $|Y,Q\rangle$ that fulfills constraints
\begin{align}
\langle Y,Q|\hat{X}_I|Y,Q\rangle = Y_I\, , 
\qquad
\langle Y,Q|\hat{P}_I|Y,Q\rangle = Q_I\, .
\end{align}
The state minimizing the energy
\begin{align}
\langle Y,Q|\hat{H}|Y,Q\rangle
\end{align}
is particularly interesting. 
It is important to note that such a $|Y,Q\rangle$ belongs to the extended Hilbert space ${\cal H}_{\rm ext}$, but not necessarily to the gauge-invariant Hilbert space ${\cal H}_{\rm inv}$. It undergoes transformation as
\begin{align}
|Y,Q\rangle\to|U^{-1}YU,U^{-1}QU\rangle
\end{align}
through gauge transformation, as illustrated in Fig.~\ref{fig:wave-packet}. States that transform into each other via gauge transformation are considered identical. Notably, the shape of the wave packet remains invariant under gauge transformation. The only ``matrices" that can be diagonalized are $Y_I$ and $Q_I$.

\begin{figure}[htbp]
  \begin{center}
   \includegraphics[width=80mm]{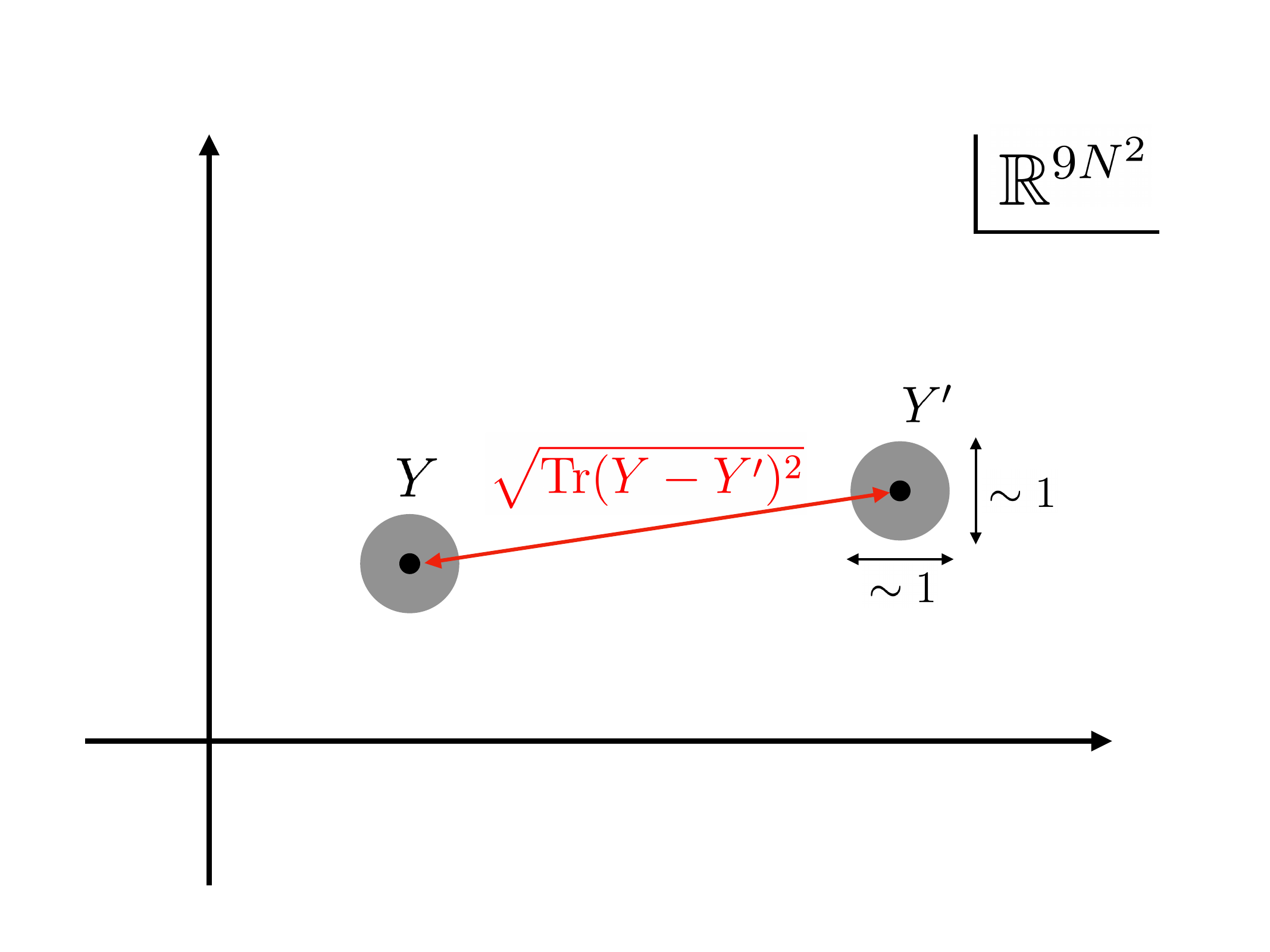}
  \end{center}
  \caption{
Each wave packet smoothly extends in $\mathbb{R}^{9N^2}$. Along each of the $9N^2$ dimensions, the size of the wave packet is at most of the order $N^0$. Two wave packets centered around $Y$ and $Y'$ in $\mathbb{R}^{9N^2}$ exhibit vanishingly small overlap if the centers of the wave packets are well separated, i.e., $\sqrt{{\rm Tr}(Y-Y')^2}\gg 1$. This figure is adapted from Ref.~\cite{Gautam:2022akq}.
  }\label{fig:wave-packet-distance}
\end{figure}

Each wave packet smoothly extends in both coordinate space $\mathbb{R}^{9N^2}$ and momentum space $\mathbb{R}^{9N^2}$. Along each of the $9N^2$ dimensions, the size is at most of order $N^0$. Two wave packets centered around $Y$ and $Y'$ in $\mathbb{R}^{9N^2}$ exhibit vanishingly small overlap if the centers of wave packets are well separated, i.e., $\sqrt{{\rm Tr}(Y-Y')^2}\gg 1$; see Fig.~\ref{fig:wave-packet-distance}.

%%%%%%%%%%%%%%%
%%%%%%%%%%%%%%%
\subsubsection*{Gaussian matrix model and coherent state}
%%%%%%%%%%%%%%%
%%%%%%%%%%%%%%%
We will examine the Gaussian matrix model as an illustrative example. We set the coupling in the toy model Hamiltonian~\eqref{eq:toy-Hamiltonian} to zero, while assuming $m$ and $\omega$ to be 1:
\begin{align}
\hat{H}
=
\sum_I
{\rm Tr}\left(
\frac{1}{2}\hat{P}_I^2
+
\frac{1}{2}\hat{X}_I^2
\right)
=
\frac{1}{2}
\sum_{I,\alpha}\left(\hat{P}_{I,\alpha}^2+\hat{X}_{I,\alpha}^2\right). 
\end{align}
Specifically, we examine $9N^2$ harmonic oscillators under the SU($N$) gauge symmetry. The ground state simply corresponds to the tensor product of $9N^2$ Fock vacua,
\begin{align}
|0\rangle
\equiv
\otimes_{I,\alpha}|0\rangle_{I,\alpha}.
\end{align}
The ground state $|0\rangle$ is SU($N$)-invariant. 
The wave packet defined above is the coherent state, 
\begin{align}
|Y,Q\rangle
=
e^{-i\sum_I{\rm Tr}(Y_I\hat{P}_I-Q_I\hat{X}_I)}|0\rangle. 
\label{coherent-state}
\end{align}
The ground state $|0\rangle$ is a wave packet about $Y=0$ and $Q=0$.
%%%%%%%%%%%%%%%
%%%%%%%%%%%%%%%
\subsubsection*{Superselection at large $N$}
%%%%%%%%%%%%%%%
%%%%%%%%%%%%%%%
The wave packet $|Y,Q\rangle$ defined above lacks SU($N$) invariance unless $Y$ and $Q$ are proportional to the identity matrix. We can interpret it as a superselection sector within the gauge-invariant Hilbert space as follows~\cite{Hanada:2019czd}. 
We can obtain the gauge-invariant counterpart of $|Y,Q\rangle$ by symmetrizing it over all possible SU($N$) transformations:
\begin{align}
\mathcal{N}^{-1/2}\int dU |U^{-1}YU,U^{-1}QU\rangle\, ,
\label{eq:gauge-symmetrization}
\end{align}
where $\mathcal{N}$ is a normalization factor.
In the large-$N$ limit, the spread of the wave function along each direction $(I,\alpha)$ is of order $N^0$ or less, while typical eigenvalues of $Y_I$ and $Q_I$ are of order $\sqrt{N}$. Even for small SU($N$) transformations (specifically, $U=e^{i\epsilon^\alpha\tau_\alpha}$, where $\sum_\alpha(\epsilon^\alpha)^2$ is small but of order $N^0$), the overlap between $|Y,Q\rangle$ and $|U^{-1}YU,U^{-1}QU\rangle$ becomes parametrically small.
For coherent states, we can explicitly observe that the overlap behaves as $e^{-\frac{1}{8}\sum_I{\rm Tr}(Y_I-U^{-1}Y_IU)^2}$ when $Q=0$. In typical situations of interest, ${\rm Tr}(Y^2)$ grows with $N$. Therefore, the overlap is essentially a delta function.

As observables, we focus on gauge-invariant operators that do not significantly change the energy of the system, such as ${\rm Tr}(\hat{X}_{I_1}\hat{X}_{I_2}\cdots\hat{X}_{I_k})$, where $k=O(N^0)$. Such operators cannot connect $|Y,Q\rangle$ and $|U^{-1}YU,U^{-1}QU\rangle$ unless $U$ is parametrically close to $G_{Y,Q}$, where $G_{Y,Q}\equiv\{g\in G| g^{-1}Yg=Y, g^{-1}Qg=Q\}$ represents the stabilizer of $(Y,Q)$.
Whether we compute the expectation value of such an operator using $|Y,Q\rangle$ or its gauge-invariant counterpart, we obtain the same value. In this sense, different gauge choices correspond to different superselection sectors. Both $|Y,Q\rangle$ and its gauge symmetrization \eqref{eq:gauge-symmetrization} produce identical expectation values, and thus, they cannot be distinguished.
%%%%%%%%%%%%%%%
%%%%%%%%%%%%%%%
\subsection{Geometry from matrix wave packets}
%%%%%%%%%%%%%%%
%%%%%%%%%%%%%%%
Matrices $Y_I$ are related to the geometry consisting of D-branes and open strings through Witten's interpretation~\cite{Witten:1995im}.
Let us list a few features justifying this identification~\cite{Hanada:2021ipb}:
\begin{itemize}
\item
Reference~\cite{Witten:1995im} examined low-energy states, which should correspond to wave packets. The only ``matrices" associated with a wave packet that can be diagonalized are $Y_I$ and $Q_I$.

\item
Let us consider a diagonal configuration, 
\begin{align}
(Y_I)_{ij}=y_{I,i}\delta_{ij},
\qquad
\vec{y}_i=(y_{1,i},\cdots,y_{9,i})\in\mathbb{R}^9\, , 
\nonumber\\
(Q_I)_{ij}=q_{I,i}\delta_{ij},
\qquad
\vec{q}_i=(q_{1,i},\cdots,q_{9,i})\in\mathbb{R}^9\, . 
\end{align} 
Around this background, fluctuations of the $(i,j)$ component have a mass proportional to $|\vec{y}_i-\vec{y}_j|$. 
This framework allows for interpreting $\vec{y}_i$ and $\vec{q}_i$ as the location and momentum of the $i$-th D0-brane, while the off-diagonal entries describe the open strings stretched between D-branes.

\item
If the locations of D-branes coincide, the symmetry of the state is enhanced. If $N_1$ D-branes sit at $\vec{y}_1=\cdots=\vec{y}_{N_1}$ (with momentum $\vec{q}_1=\cdots=\vec{q}_{N_1}$), while another $N_2$ D-branes sit at $\vec{y}_{N_1+1}=\cdots=\vec{y}_{N_1+N_2}$ (with momentum $\vec{q}_{N_1+1}=\cdots=\vec{q}_{N_1+N_2}$), and so forth, the symmetry becomes ${\rm U}(N_1)\times{\rm U}(N_2)\times\cdots$. This is the correct symmetry enhancement. 

\end{itemize}
%%%%%%%%%%%%%%%
%%%%%%%%%%%%%%%
\subsubsection*{Partial deconfinement and small black hole}
%%%%%%%%%%%%%%%
%%%%%%%%%%%%%%%
Partially-deconfined states~\cite{Hanada:2016pwv,Hanada:2018zxn,Hanada:2020uvt,Hanada:2019czd,Berenstein:2018lrm} can describe a small black hole.
In partially-deconfined states, an SU($M$) subgroup of SU($N$) is deconfined, where $M$ depends on the energy.  
Specifically, we consider wave packets of the form
\begin{align}
Y_I^{[i]}
=
\left(
\begin{array}{cc}
 \tilde{Y}_I^{[i]}& 0\\
0& 0
\end{array}
\right), 
\qquad
Q_I^{[i]}
=
\left(
\begin{array}{cc}
 \tilde{Q}_I^{[i]}& 0\\
0& 0
\end{array}
\right)\,  
\label{partially-deconfined-state}
\end{align}
where $i=1,2,\cdots$ labels different wave packets in this class,
and their linear combinations, 
\begin{align}
\sum_{i}c_i\ket{Y^{[i]},Q^{[i]}}\, . 
\end{align}
 We assume $\tilde{Y}_I^{[i]}$ and $\tilde{Q}_I^{[i]}$ are significantly excited, e.g., ${\rm Tr}(\tilde{Y}_I^{[i]})^2\sim {\rm Tr}(\tilde{Q}_I^{[i]})^2\sim M^2$, while the other blocks are set to zero. Both $M$ D-branes and open strings between them are excited, forming an extended object. The remaining $N-M$ D-branes are positioned at the origin with no open strings excited between them or between those $N-M$ D-branes and the other $M$ D-branes. It is important to note that all D-branes contribute to the exterior geometry. 
 
When $M\ll N$, the coarse-grained entropy of the partially-deconfined phase, which should agree with the Bekenstein-Hawking entropy when there is a weakly-curved gravity dual, can be obtained from a typical partially-deconfined state as von Neumann entropy by tracing out the confined sector~\cite{Gautam:2022akq}. 

%%%%%%%%%%%%
%%%%%%%%%%%%
\subsection{Gravity dual}
\hspace{0.51cm}
%%%%%%%%%%%%
%%%%%%%%%%%%
The embedding of geometry into matrices discussed above did not assume the details of the models. 
Whether weakly-coupled dual gravitational description exists depend on the details. 
Supersymmetric matrix models, specifically the BFSS model~\cite{deWit:1988wri,Banks:1996vh,Itzhaki:1998dd} and the BMN model~\cite{Berenstein:2002jq} provide well-controlled gravity duals that have been tested quantitatively via numerical simulations, starting with Refs.~\cite{Anagnostopoulos:2007fw,Catterall:2008yz}. See Ref.~\cite{Bergner:2021goh} and Ref.~\cite{Pateloudis:2022ijr} for recent investigation in the context of type IIA superstring theory and M-theory, respectively. 

Seen from matrix model, the appearance of the causal structure of the bulk geometry, specifically the speed of light, is highly nontrivial. In this context, Ref.~\cite{Kabat:1999yq} discussed that, if `D0-brane' (wave packet in matrix model that corresponds to D0-brane) moves faster than the speed of light in the bulk geometry, tachyonic modes appear. This could mean either D0-brane dynamically slows down or the interpretation as semiclassical geometry breaks down. 

%%%%%%%%%%%%
%%%%%%%%%%%%
\section{Algebra of small excitations in matrix model}\label{sec:matrix_model}
\hspace{0.51cm}
%%%%%%%%%%%%
%%%%%%%%%%%%
In this section, we use the matrix wave packet reviewed in Sec.~\ref{sec:matrix-geometry-review} to probe an arbitrary region of the emergent geometry in the matrix model. 
%%%%%%%%%%%%
%%%%%%%%%%%%
\subsection{Small excitations around the ground state}
\begin{figure}[htbp]
\begin{center}
\scalebox{0.2}{
\includegraphics{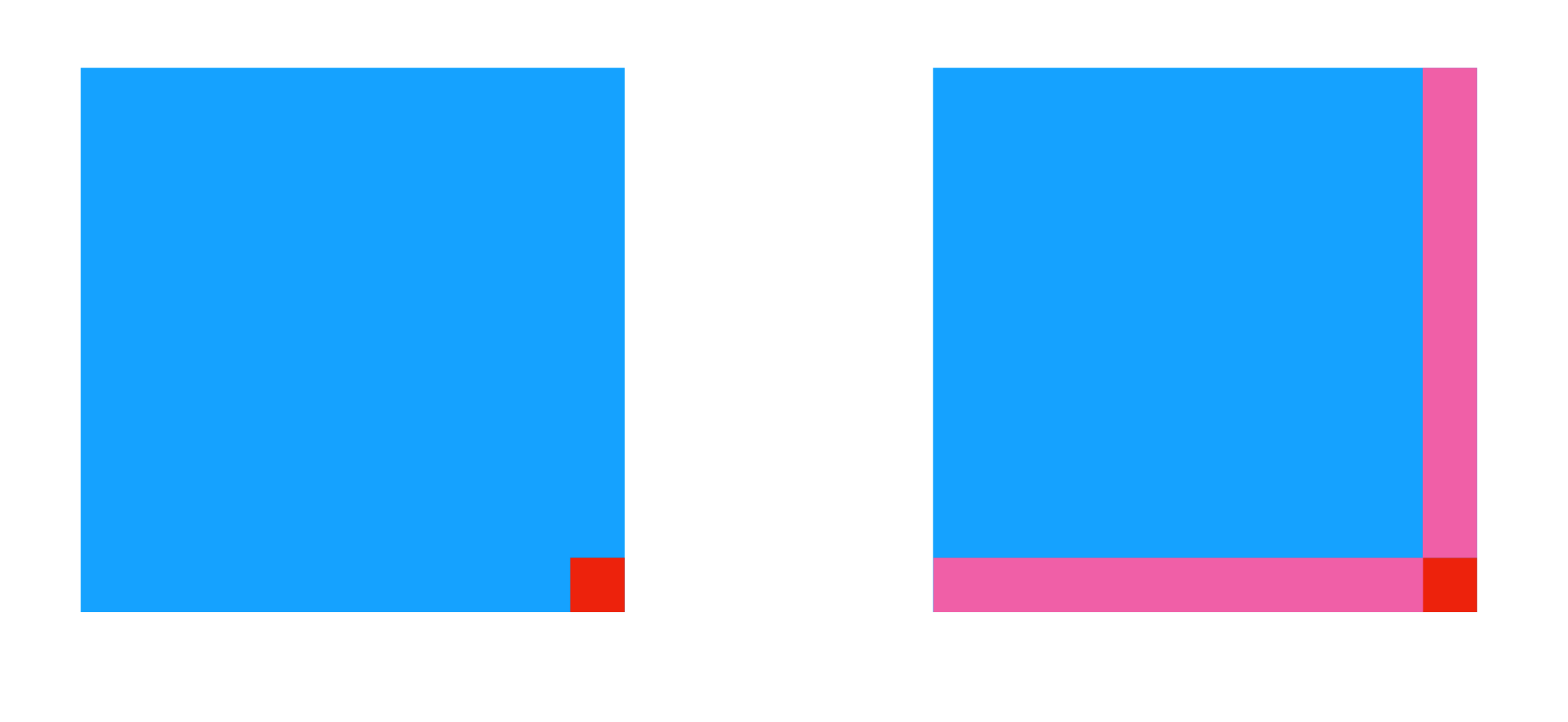}}
\end{center}
\caption{One object is excited (red). [Left] free theory. Off-diagonal blocks are not affected. [Right] Interacting theory. Off-diagonal entries (pink) are affected; specifically, though the center remains zero, width of quantum fluctuation changes. 
Such effects are small when the excited block does not go too far from the origin of $\mathbb{R}^9$. 
}\label{matrix-1-excitation}
\end{figure}

\hspace{0.51cm}
%%%%%%%%%%%%
%%%%%%%%%%%%
Firstly, we consider small excitations around the ground state.\footnote{
For the BFSS matrix model, we consider the threshold bound state~\cite{Sethi:1997pa,Yi:1997eg,Frohlich:1999zf}.
} The ground state is a wave packet in $\mathbb{R}^{9N^2}$ localized around $Y_I=0$ in the coordinate basis and $Q_I=0$ in the momentum basis. 
To excite a small extended object, we can take a wave packet around $N\times N$ matrices $Y_I$ and $Q_I$ that are nonzero only in the lower-right $n\times n$ block where $n$ is an order $N^0$ number (see Fig.~\ref{matrix-1-excitation}): 
\begin{align}
Y_I
=
\left(
\begin{array}{cc}
0 & 0\\
0 & \tilde{Y}_I
\end{array}
\right)\, , 
\qquad
Q_I
=
\left(
\begin{array}{cc}
0 & 0\\
0 & \tilde{Q}_I
\end{array}
\right)\, , 
\end{align}
where
\begin{align}
\tilde{Y}_I
=
y_I\cdot\text{1}_n
+
\tilde{\tilde{Y}}_I\, , 
\qquad
\tilde{Q}_I
=
q_I\cdot\text{1}_n
+
\tilde{\tilde{Q}}_I\, .  
\end{align}
Here, $y_I$ and $q_I$ are the location and momentum of the extended object and they can be of order $\sqrt{N}$.  
$\vec{y}=(y_1,\cdots,y_9)\in\mathbb{R}^9$ and $\vec{q}=(q_1,\cdots,q_9)\in\mathbb{R}^9$ specify the location and momentum of the object in the nine-dimensional target space of string theory. 
$\tilde{\tilde{Y}}_I$ and $\tilde{\tilde{Q}}_I$ are traceless and take at most order $N^0$ values. 
In the Gaussian limit, such a wave packet is a coherent state \eqref{coherent-state} (the left panel in Fig.~\ref{matrix-1-excitation}). 
%\begin{align}
%\ket{Y,Q}
%=
%e^{-i\mathrm{Tr}(Y_I\hat{P}_I-Q_I\hat{X}_I)}
%\ket{Y=0,Q=0}
%\end{align}
%where $\ket{Y=0,Q=0}$ is the Fock vacuum (the left panel in Fig.~\ref{matrix-1-excitation}). 
We can make the gauge-invariant version of this operator as
\begin{align}
\hat{\mathcal{O}}_{Y,Q}
=
\frac{1}{\mathrm{vol}(\mathrm{SU}(N))}
\int_{\mathrm{SU}(N)} dg\ e^{-i\mathrm{Tr}((gY_Ig^{-1})\hat{P}_I-(gQ_Ig^{-1})\hat{X}_I)}\, . 
\label{eq:SU(N)-symmetrization}
\end{align}
Note that, by construction, $\hat{\mathcal{O}}_{Y,Q}=\hat{\mathcal{O}}_{hYh^{-1},hQh^{-1}}$ for any $h\in\mathrm{SU}(N)$. 

When there is an interaction, the off-diagonal blocks (last $n$ rows and columns on the right panel in Fig.~\ref{matrix-1-excitation} shown in pink) are also affected. Specifically, as the eigenvalues of $\tilde{Y}_I$ become large, the width of the quantum fluctuation of the off-diagonal entry shrinks. This can easily be understood by noticing that, when $|\vec{y}|$ is large, off-diagonal blocks behave as harmonic oscillators whose frequency is proportional to $|\vec{y}|$. A conceptually simple way to obtain such a low-energy wave packet is to minimize the energy $\bra{Y,Q}\hat{H}\ket{Y,Q}$ fixing the center of the wave packet to $\bra{Y,Q}\hat{X}_{I,ij}\ket{Y,Q}=Y_{I,ij}$ and $\bra{Y,Q}\hat{P}_{I,ij}\ket{Y,Q}=Q_{I,ij}$. Note that the back-reaction to the $(N-n)\times(N-n)$ block can be neglected in the large-$N$ limit with fixed $n$.

\begin{figure}[htbp]
\begin{center}
\scalebox{0.2}{
\includegraphics{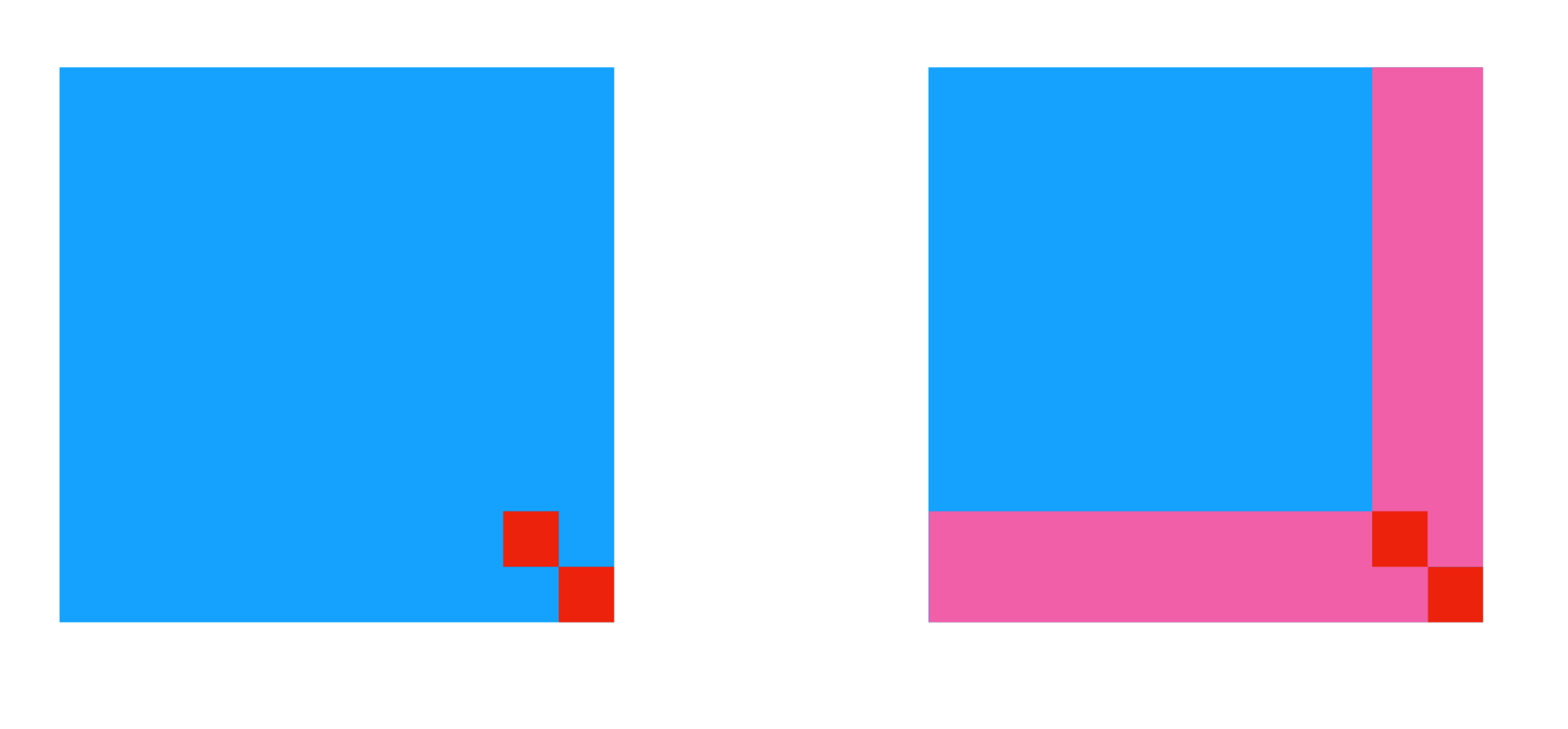}}
\end{center}
\caption{Two objects are excited (red). [Left] free theory. Off-diagonal blocks are not affected. [Right] Interacting theory. Off-diagonal entries (pink) can be affected. (See the caption of Fig.~\ref{matrix-1-excitation}.)
}\label{matrix-2-excitations}
\end{figure}

We can act two or more operators $\hat{\mathcal{O}}_{Y,Q}$, $\hat{\mathcal{O}}_{Y',Q'}$, $\cdots$ (see Fig.~\ref{matrix-2-excitations}). In the large-$N$ limit, we can neglect the possibility of exciting the same color degrees of freedom. If we act a finite number of such operators, say $k$ of them, and send $N$ to infinity, then $k$ independent blocks are excited.
For example, for $k=2$, we can write it as
\begin{align}
\hat{\mathcal{O}}_{Y,Q}\hat{\mathcal{O}}_{Y',Q'}
=
\hat{\mathcal{O}}_{Y\oplus Y',Q\oplus Q'}\, , 
\end{align}
where 
\begin{align}
Y\oplus Y'
\equiv
\left(
\begin{array}{ccc}
0 & 0 & 0\\
0 & \tilde{Y}_I & 0\\
0 & 0 & \tilde{Y}'_I
\end{array}
\right)\, , 
\qquad
Q\oplus Q'
\equiv
\left(
\begin{array}{ccc}
0 & 0 & 0\\
0 & \tilde{Q}_I & 0\\
0 & 0 & \tilde{Q}'_I
\end{array}
\right)\, . 
\end{align}
Generalizations to generic $k$ are straightforward. 
Interaction between small excitations is higher order in the $1/N$-expansion. 
%That we neglect such interactions means we consider a generalized free field theory.

By collecting the operators $\hat{\mathcal{O}}_{Y,Q}$ whose centers $\vec{y}=(y_1,\cdots,y_9)$ are in a region $R\subset\mathbb{R}^9$ and taking product of them, we can construct \textit{operator algebra corresponding to region $R$}.

For practical applications, it would be convenient to consider a state obtained by acting a gauge-invariant operator consisting of $n$ matrices, e.g., $\mathrm{Tr}(\hat{X}_{I_1}\hat{X}_{I_2}\cdots\hat{X}_{I_n})$, to the ground state. Such an operator excites an $n\times n$ block, because   
\begin{align}
\mathrm{Tr}(\hat{X}_{I_1}\hat{X}_{I_2}\cdots\hat{X}_{I_n})=\hat{X}_{I_1,12}\hat{X}_{I_2,23}\cdots\hat{X}_{I_n,n1}+\textrm{permutations}\, 
\label{eq:excitation_block_size}
\end{align}
up to $1/N$-suppressed terms. 

%%%%%%%%%%%%
%%%%%%%%%%%%
\subsection{Small excitation around black hole geometry}
\begin{figure}[htbp]
\begin{center}
\scalebox{0.2}{
\includegraphics{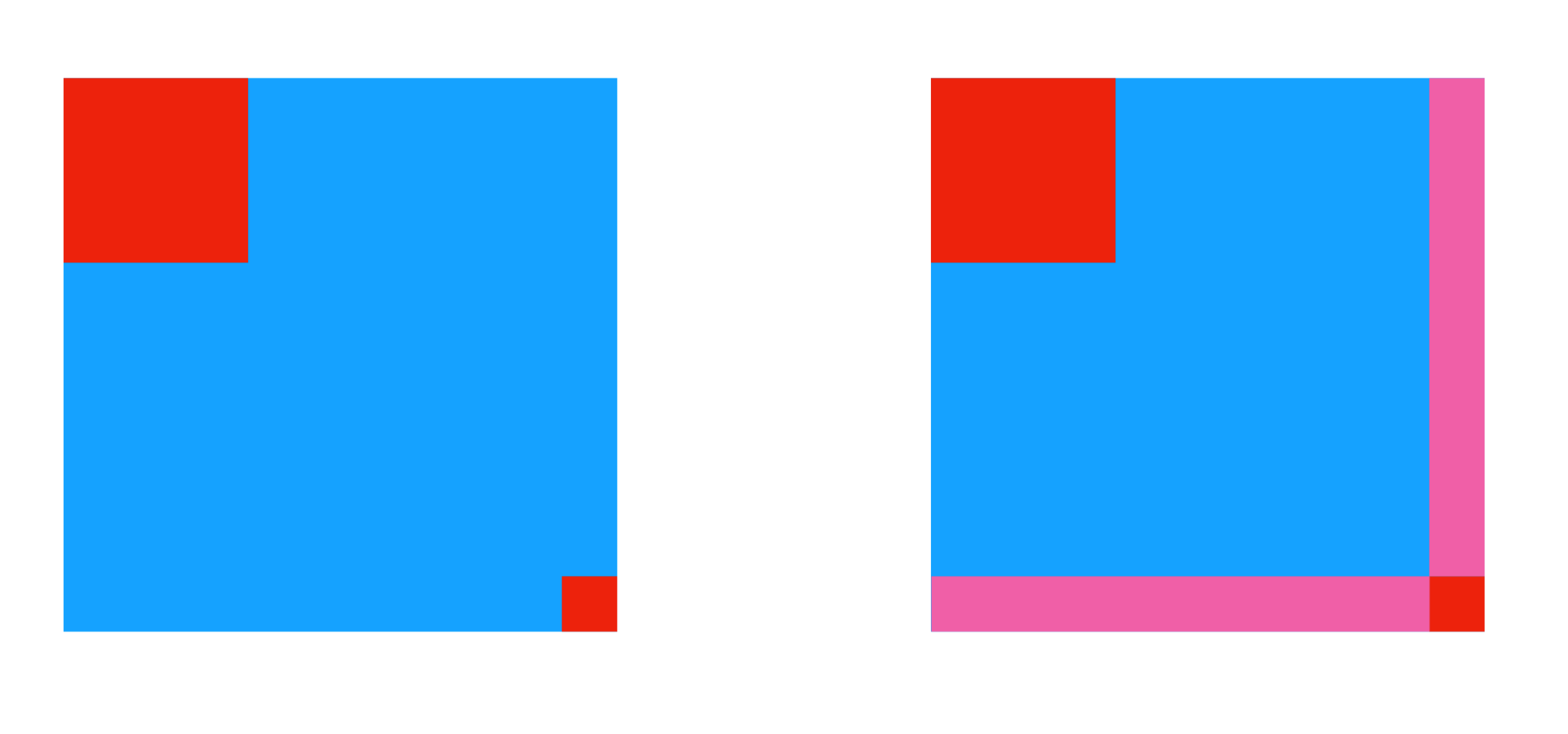}}
\end{center}
\caption{One object is excited (small red block) on top of a partially-deconfined state. The large red block is the SU($M$)-deconfined sector. [Left] free theory. Off-diagonal blocks are not affected. [Right] Interacting theory. Off-diagonal entries (pink) can be affected. (See the caption of Fig.~\ref{matrix-1-excitation}.)}\label{matrix-1-excitation-PD}
\end{figure}

\hspace{0.51cm}
%%%%%%%%%%%%
%%%%%%%%%%%%
Next, we consider an SU($M$)-partially-deconfined state. We can prepare either a pure state or a mixed state. (It is conjectured that the 11d BH is described by the partially-deconfined states in the matrix model~\cite{Bergner:2021goh}.) If small excitations are well separated from the partially-deconfined sector (specifically, if $\vec{y}$ is larger than the eigenvalues of the SU($M$)-deconfined block), we can repeat essentially the same logic. The only difference is that we should take a superselection sector, e.g., the deconfined sector sits at the upper-left $M\times M$ block as depicted in Fig.~\ref{matrix-1-excitation-PD}~\cite{Hanada:2019czd}, and define the operators that create excitations in the confined sector.
Fixing the superselection sector on the gauge theory side in this way corresponds to fixing the semiclassical geometry on the gravity side~\cite{Leutheusser:2022bgi}. 

Because a matrix wave packet has the information of locations and momenta of small excitations, 
we can build both `algebra of states that stay outside BH' (such excitations can be created in supersymmetric theories due to the flat directions), and `algebra of states that fall into BH'. 
Late-time behaviors of operators in the former and latter algebras must be drastically different. See Ref.~\cite{Gesteau:2024dhj} for the relationship between the late-time behaviors and types of algebras. 

%%%%%%%%%%%%
%%%%%%%%%%%%
\subsection{A remark on terminology}\label{sec:terminology}
\hspace{0.51cm}
%%%%%%%%%%%%
%%%%%%%%%%%%
A remark is in order concerning the nuance of the word \textit{matrix wave packet}.
As we have seen, it is of crucial importance to consider wave packets in the space of matrix degrees of freedom, which spreads in $\mathrm{R}^{9N^2}$ in the case of the D0-brane matrix model. Naturally, we called such a wave packet in $\mathrm{R}^{9N^2}$ \textit{matrix wave packet}. 
In the rest of the paper, we will mainly consider small excitations added to semi-classical geometry. Specifically, we are interested in objects expressed by small blocks in $X_I$ corresponding to the nonzero part of $Y_I$ and $Q_I$ (e.g., small red blocks in Fig.~\ref{matrix-1-excitation}, Fig.~\ref{matrix-2-excitations}, and Fig.~\ref{matrix-1-excitation-PD}). Therefore, we will use the same word \textit{matrix wave packet} to mean these small blocks as well. Ultimately, the difference is only in their nuances --- whether our focus is on the entire geometry or small excitations --- because the full $N\times N$ matrices contain all sub-blocks.

%%%%%%%%%%%%
%%%%%%%%%%%%
\section{QFT wave packet and AdS$_5$/CFT$_4$ duality}\label{sec:QFT}
\hspace{0.51cm}
%%%%%%%%%%%%
%%%%%%%%%%%%
In this section, we discuss emergent geometry in QFT, specifically 4d $\mathcal{N}=4$ super Yang-Mills on S$^3$. (Essentially the same picture holds on $\mathbb{R}^3$. For simplicity, we use $\mathbb{R}^3$ for concrete computations. The same expressions are valid for S$^3$ near the boundary.) We call the three-dimensional space of QFT as QFT space, in order to distinguish it from the space that emerges from matrix degrees of freedom. This theory has six scalars and their momentum conjugates.  
We denote them as $\hat{X}_{I,ij,\vec{x}}$ and $\hat{P}_{I,ij,\vec{x}}$, where $I$ runs from 1 to 6, $\vec{x}$ is the coordinate in the QFT space, and $i,j=1,\cdots,N$.
The canonical commutation relation is
\begin{align}
[\hat{X}_{I,ij,\vec{x}},\hat{P}_{J,kl\vec{x}'}]
=
i\delta_{IJ}\delta_{il}\delta_{jk}\delta^3(\vec{x}-\vec{x}')\, . 
\end{align}

There are at least two kinds of wave packets: a wave packet in the QFT space which we call \textit{QFT wave packet}, and a wave packet in the space of matrix degrees of freedom which we call \textit{matrix wave packet}. A QFT wave packet can specify the location in the three-dimensional space. A matrix wave packet can specify the location in $\mathbb{R}^6$. AdS$_5$ is described by four-dimensional spacetime in QFT and the radial coordinate of $\mathbb{R}^6$ from matrices. S$^5$ is described by the angular coordinates of $\mathbb{R}^6$. 
We use the name \textit{bulk wave packet} for an excitation which is a QFT wave packet and matrix wave packet simultaneously.
A location in AdS$_5\times$S$^5$ can be associated with a bulk wave packet. We argue that a bulk wave packet describes a closed string.

%%%%%%%%%%%%%%%
%%%%%%%%%%%%%%%
\subsection{Global matrix wave packet and D3-brane in AdS$_5$/CFT$_4$}\label{sec:D3-brane}
%%%%%%%%%%%%%%%
%%%%%%%%%%%%%%%
Let us start with a review of the matrix wave packet in QFT. This is a straightforward generalization of Sec.~\ref{sec:matrix-geometry-review}.
The partition function is the same as \eqref{eq:Z-H-ext-MM}, except that we need to consider all possible local gauge transformations instead of $G$ that can be schematically written as $\mathcal{G}=\otimes_{\vec{x}}G_{\vec{x}}$ where $G_{\vec{x}}$ is the gauge group $G$ acting on a spatial point $\vec{x}$ in the QFT space. We can use a lattice regularization to make sense of this expression. The canonical partition function can be written as 
\begin{align}
Z(T)
&=
\frac{1}{{\rm vol}\mathcal{G}}\int_\mathcal{G}dg
{\rm Tr}_{{\cal H}_{\rm ext}}\left(
\hat{g}
e^{-\hat{H}/T}
\right). 
\label{eq:Z-H-ext-QFT}
\end{align}

\begin{figure}[htbp]
\begin{center}
\scalebox{0.2}{
\includegraphics{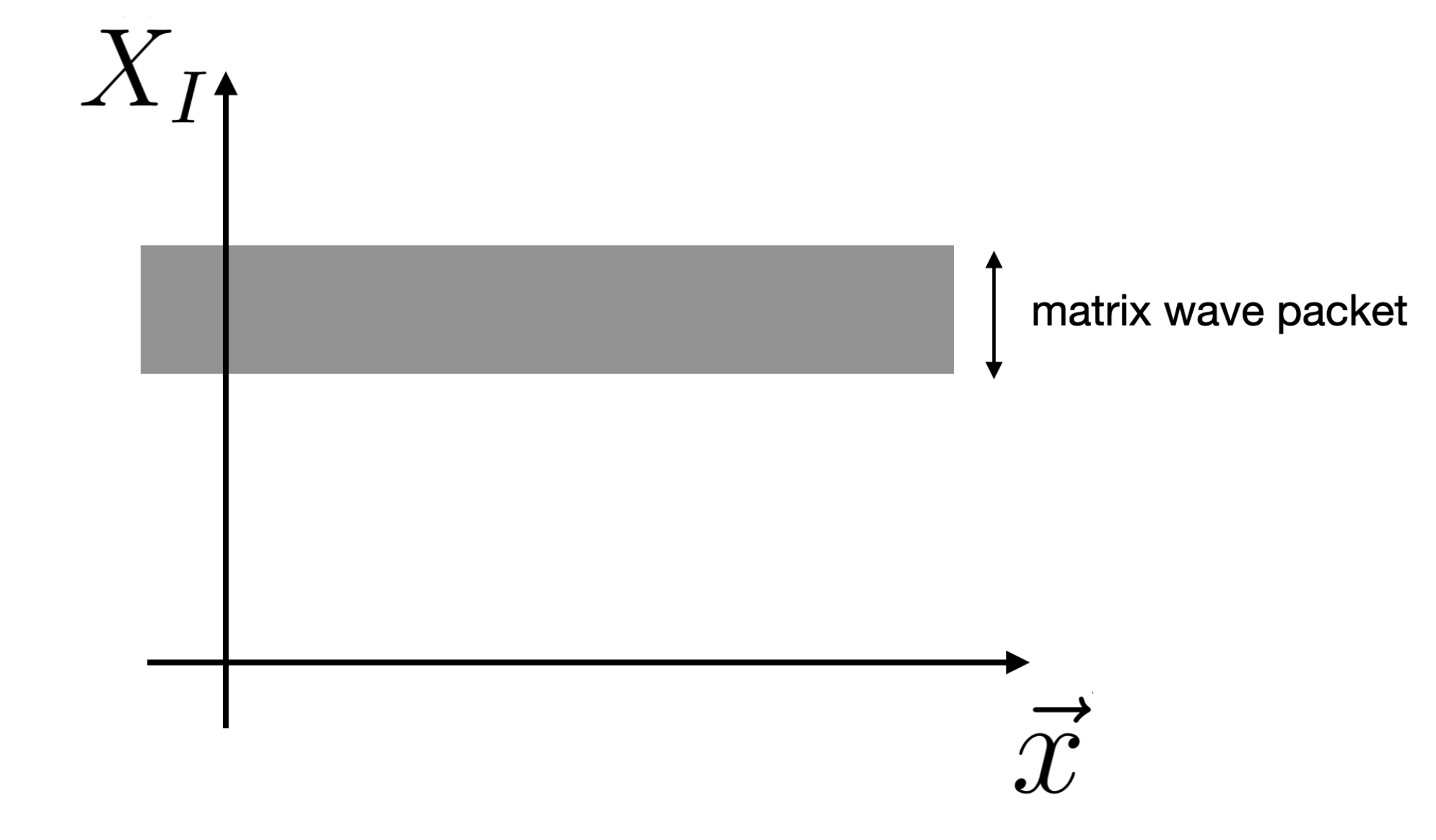}}
\end{center}
\caption{A schematic picture of a global matrix wave packet, e.g., a D3-brane. 
It requires excitation at each spatial point $\vec{x}$. 
More precisely, small sub-blocks of scalar matrices describe D3-branes (see Sec. 3.1.3 for a remark on terminology).
}\label{matrix-wave-packet-in-QFT}
\end{figure}

As a concrete example, let us consider $(3+1)$-dimensional maximal super Yang-Mills theory. 
The D3-brane geometry on $\mathbb{R}^6=\mathbb{R}_{>0}\times {\rm S}^5$ can be generated from six scalar fields. $\mathbb{R}_{>0}$ is interpreted as the radial coordinate of AdS$_5$~\cite{Maldacena:1997re}. 
If we use a lattice regularization for the spatial dimensions, wave functions are defined on $\mathbb{R}^{6\times N^2\times n_{\rm site}}$, where $6$ is the number of scalar fields and $n_{\rm site}$ is the number of spatial lattice sites. The center of a wave packet is specified by $Y_I(\vec{x})$ and $Q_I(\vec{x})$. Note that we assume $\vec{x}$ runs through the 3-dimensional space of QFT, and we assume they are continuous functions in the continuum limit. (Strictly speaking, we have to take into account the spatial directions of the gauge field as well, whose discretized version lives on the links.) 
In the ground state,\footnote{This is the case for the theory on S$^3$. On $\mathbb{R}^3$, $Y_I=0$, $Q_I=0$ is one of the super-selection sectors.} $Y_I=0$ and $Q_I=0$, and hence, all D3-branes are sitting at the origin of AdS$_5$ and no open string is excited. 
This is exactly the expected property of the BPS limit of the black three-brane. 
We can take some D3-branes out of the BPS black hole by giving nontrivial values to $Y_I$ and $Q_I$. This is essentially the Higgsing of scalar fields. See Fig.~\ref{matrix-wave-packet-in-QFT}. 

A quantum state describing a D3-brane with a nontrivial shape can be written as 
\begin{align}
\int_{\vec{x}\in\mathrm{S}^3} [dX_{\vec{x}}]\ F[X_{\vec{x}}]\ket{X_{\vec{x}}}\, , 
\label{eq:matrix_wave_packet_S3}
\end{align}
where $F[X_{\vec{x}}]$ is localized around a certain value $Y_I(\vec{x})$ at each $\vec{x}$. 
Specifically, we can observe excitations at different spatial points in S$^3$ simultaneously, corresponding to the location of the D3-brane away from $\vec{0}\in\mathbb{R}^6$. We use a name \textit{global matrix wave packet} to denote such a matrix wave packet. Here, `global' means it spreads globally on S$^3$. 
An important point for holographic duality is:
\begin{itemize}
\item
Global matrix wave packet describes D3-brane (black 3-brane). 
\end{itemize}
More precisely, small sub-blocks of scalar matrices describe D3-branes (see Sec.~\ref{sec:terminology} for a remark on terminology).
In this context, the values of scalars specify the locations of the D3-branes in the bulk (specifically, S$^5$ and the radial coordinate of AdS$_5$). This can be seen through the effective description of Higgsed phase: the Dirac-Born-Infeld action can be obtained~\cite{Maldacena:1997re}, and further nontrivial evidence is obtained from the consistency of special conformal transformation on the gravity side and gauge theory side~\cite{Jevicki:1998qs}.
The same are expected for D$p$-brane and $(p+1)$-dimensional super Yang-Mills theory for $p\neq 3$ as well~\cite{Maldacena:1997re,Jevicki:1998ub}. 

%In Sec.~\ref{sec:QFT-and-bulk-wave-packets}, we will also consider another kind of matrix wave packet, which we call bulk wave packet.  

%%%%%%%%%%%%
%%%%%%%%%%%%
\subsection{QFT wave packet and bulk wave packet}\label{sec:QFT-and-bulk-wave-packets}
\hspace{0.51cm}
%%%%%%%%%%%%
%%%%%%%%%%%%
Next, let us introduce QFT wave packets and bulk wave packets. 
To simplify the notation and see the essence, let us use one-component real scalar $\hat{\phi}_{\vec{x}}$ and its momentum conjugate $\hat{\pi}_{\vec{x}}$. For each $I$ and $\alpha$, $\hat{X}_{I,\vec{x}}^\alpha$ and $\hat{P}_{I,\vec{x}}^\alpha$ behaves similar to $\hat{\phi}_{\vec{x}}$ and $\hat{\pi}_{\vec{x}}$. We keep the spatial dimension $d$ general in this subsection, though we will focus on $d=3$ later. 
The canonical commutation relation is
\begin{align}
[\hat{\phi}_{\vec{x}},\hat{\pi}_{\vec{x}'}]
=
i\delta^d(\vec{x}-\vec{x}')\, . 
\end{align}

We can make smeared operators that can create wave packets in the QFT space, which we call \textit{QFT wave packets}. A simple example is the wave packets created by Gaussian-smeared field variables
\begin{align}
\hat{\Phi}_{\vec{x}_0,\vec{p}_0}
&\equiv
\frac{1}{(2\pi\sigma^2)^{d/2}}
\int d^d\vec{x}e^{-\frac{(\vec{x}-\vec{x}_0)^2}{2\sigma^2}-i\vec{p}_0\cdot(\vec{x}-\vec{x}_0)}\hat{\phi}_{\vec{x}}\, , 
\nonumber\\
\hat{\Pi}_{\vec{x}_0,\vec{p}_0}
&\equiv
\frac{1}{(2\pi\sigma^2)^{d/2}}
\int d^d\vec{x}e^{-\frac{(\vec{x}-\vec{x}_0)^2}{2\sigma^2}-i\vec{p}_0\cdot(\vec{x}-\vec{x}_0)}\hat{\pi}_{\vec{x}}\, . 
\label{eq:Gaussian_smeared_operators}
\end{align}
These operators can create wave packets in the QFT space, localized around $\vec{x}_0$ in the coordinate space and $\vec{p}_0$ in the momentum space. The commutation relation is
\begin{align}
[\hat{\Phi}_{\vec{x}_0,\vec{p}_0},\hat{\Pi}_{\vec{x}'_0,\vec{p}'_0}^\dagger]
=
[\hat{\Phi}_{\vec{x}_0,\vec{p}_0},\hat{\Pi}_{\vec{x}'_0,-\vec{p}'_0}]
=
\frac{i}{(4\pi\sigma^2)^{d/2}}
\cdot 
e^{-\frac{(\vec{x}_0-\vec{x}'_0)^2}{4\sigma^2}-\frac{\sigma^2(\vec{p}_0-\vec{p}'_0)^2}{4}+i\frac{\vec{p}_0+\vec{p}'_0}{2}\cdot(\vec{x}_0-\vec{x}'_0)}\, . 
\label{eq:Gaussian-WP-commutator}
\end{align}

A QFT wave packet including a bulk wave packet is written schematically as 
\begin{align}
\int d^3\vec{x}f(\vec{x})\hat{\phi}_{\vec{x}}\ket{\rm g.s.}\, . 
\label{eq:smeared_phi}
\end{align}
We can also use a composite operator $\hat{O}_{\vec{x}}$ made of $\hat{\phi}_{\vec{x}}$, and we can act with multiple smeared fields on the ground state. In \eqref{eq:smeared_phi}, $\hat{\phi}_{\vec{x}}\ket{\rm g.s.}$ can be seen as a small excitation around point $\vec{x}$. We take a linear combination of such states. Therefore, unlike the global matrix wave packet, we observe significant difference from the ground state only locally. 

\begin{figure}[htbp]
\begin{center}
\scalebox{0.2}{
\includegraphics{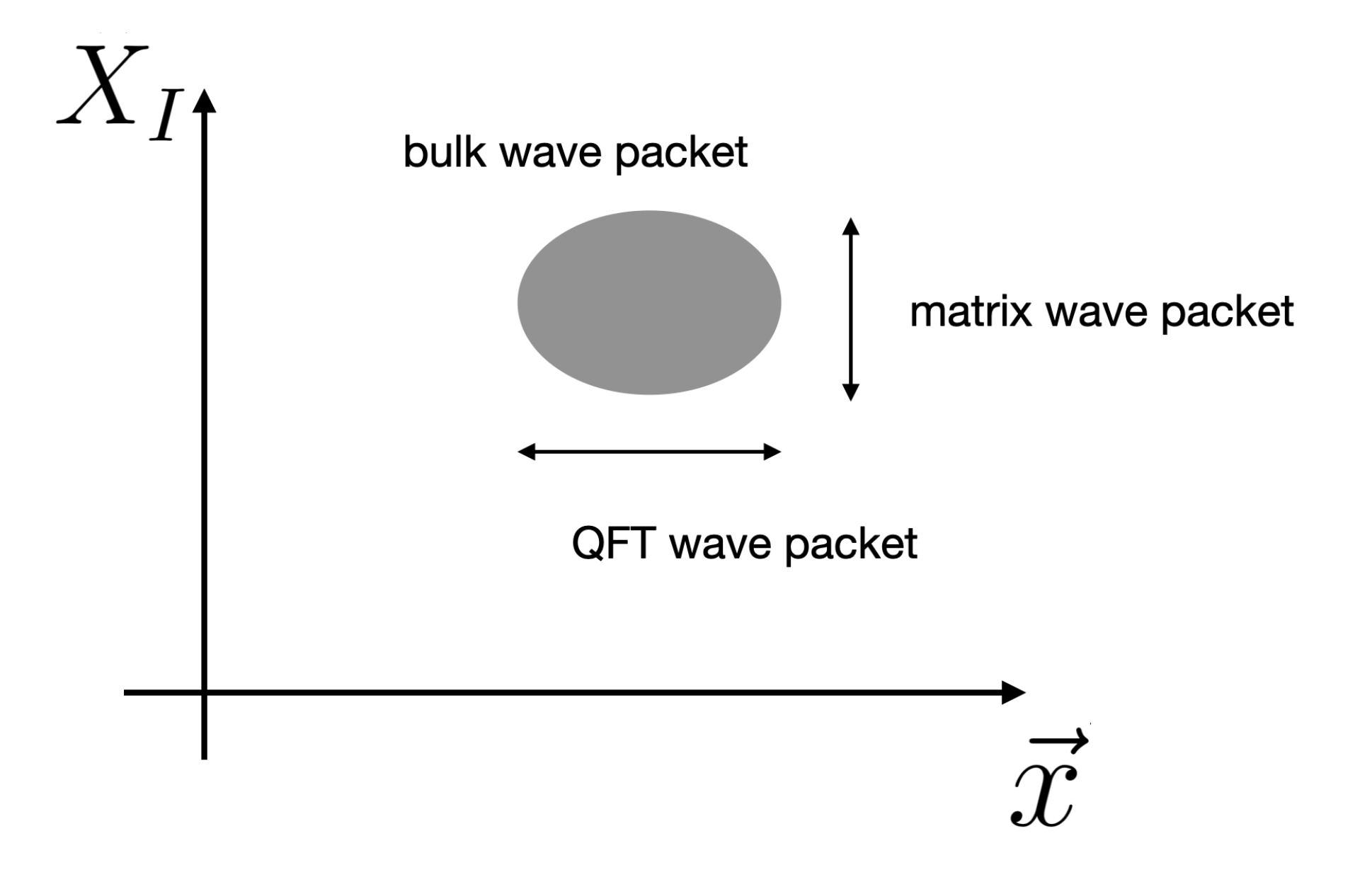}}
\end{center}
\caption{A schematic picture of a bulk wave packet in QFT (specifically, 4d $\mathcal{N}=4$ SYM).
It is obtained by smearing the locations of finitely many excitations. 
}\label{bulk_wave_packet}
\end{figure}

\begin{figure}[htbp]
\begin{center}
\scalebox{0.2}{
\includegraphics{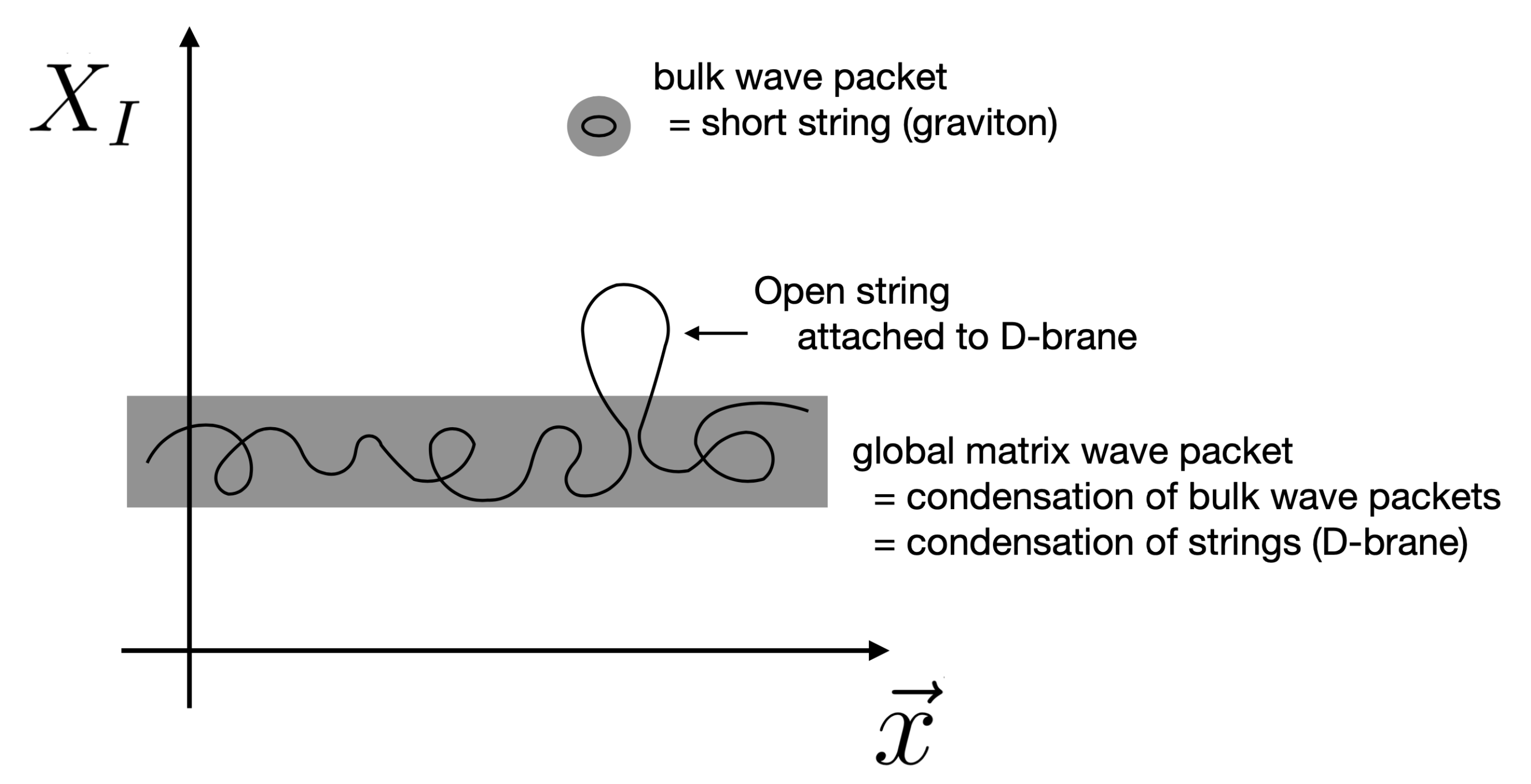}}
\end{center}
\caption{Physical interpretation of bulk wave packet and global matrix wave packet on the gravity side. Bulk wave packet describes short strings such as graviton. A global matrix wave packet is a D-brane. Note that a D-brane is a condensation of strings and a global matrix wave packet is a condensation of bulk wave packets, consistent with this interpretation. 
}\label{WP-vs-string}
\end{figure}

A QFT wave packet can be a matrix wave packet as well. We call such a wave packet that is simultaneously a QFT wave packet and matrix wave packet as \textit{bulk wave packet}; see Fig.~\ref{bulk_wave_packet}. 
We can immediately see that: 
\begin{itemize}
\item
A global matrix wave packet is a condensation of bulk wave packets; see Fig.~\ref{WP-vs-string}. 
\end{itemize}
This simple observation sits at the heart of geometric interpretation provided in Sec.~\ref{sec:geometry-bulk-wave-packet}.

%%%%%%%%%%%%%%%%%
%%%%%%%%%%%%%%%%%
\subsection{Bulk wave packet and closed string in AdS$_5$/CFT$_4$}\label{sec:geometry-bulk-wave-packet}
%%%%%%%%%%%%%%%%%
%%%%%%%%%%%%%%%%%
Now we focus on the 4d $\mathcal{N}=4$ super Yang-Mills theory and AdS$_5$/CFT$_4$ duality. 
The meaning of bulk wave packet on the gravity side can be inferred from the following:
\begin{itemize}
\item
Global matrix wave packet describes D3-brane. The values of scalars specify the location of the D3-brane in the bulk (specifically, S$^5$ and the radial coordinate of AdS$_5$)~\cite{Maldacena:1997re}.
\item
A global matrix wave packet is a condensation of bulk wave packets. 
\item
D3-brane (black 3-brane) is a condensation of strings.
\end{itemize}
Naturally, we interpret a bulk wave packet as a string or a few strings. With this identification, 
because a D3-brane ($=$ global wave packet) is a condensation of strings ($=$ bulk wave packet), 
and because the values of scalars in the global matrix wave packet specify the location of D3-brane in the bulk, 
we can naturally interpret that the values of scalars in a bulk wave packet specify the location of the corresponding string in the bulk; see Fig.~\ref{WP-vs-string}.

Below, we argue further that nontrivial aspects of other bulk reconstruction programs naturally follow from this interpretation. 
Specifically:
\begin{itemize}
\item
The size of the region on the boundary where the bulk wave packet has support is related to the radial coordinate of AdS.

\item
Quantum entanglement plays an important role in bulk reconstruction. 

\end{itemize}
Because the size of the wave packet is inversely related to the energy scale of the excitations, the correspondence between the radial coordinate and the energy scale naturally follows. In particular, when the QFT is ultraviolet-regularized e.g., by using a lattice, because the UV cutoff of the QFT $\Lambda_{\rm UV}$ limits the size of the wave packet at $\lesssim 1/\Lambda_{\rm UV}$, it also specifies the radial coordinate of the bulk excitations at $r\lesssim\Lambda_{\rm UV}$, where $r=0$ and $r=\infty$ are the center and the boundary of AdS, respectively.

We will first provide a consistency check of our interpretation of the radial coordinate in Sec.~\ref{sec:radial_coordinate} assuming the standard AdS/CFT dictionary. In Sec.~\ref{sec:radial-coordinate-qualitative}, we give a qualitative argument for the relationship between the size of the region on boundary and radial coordinate without assuming dual gravity description. We comment on AdS$_4$/CFT$_3$ correspondence in Sec.~\ref{sec:ABJM}. In Sec.~\ref{sec:geometry_from_entanglement}, it is explained how the bulk wave packet is related to quantum entanglement. In Sec.~\ref{sec:traversable-wormhole}, the similarity to the case of traversable wormhole is discussed. The entanglement associated with the bulk wave packet is O($N^0$) correction to the leading O($N^2$) part captured by the Ryu-Takayanagi formula~\cite{Ryu:2006bv} and its variants. The importance of the leading part is suggested in Sec.~\ref{sec:traversable-wormhole}. 

As a disclaimer, we comment that, although the boundaries of the bulk regions shown in Fig.~\ref{wave-packet-QFT}, Fig.~\ref{entanglement}, and Fig.~\ref{entanglement_disjoint_regions} look like Ryu-Takayanagi (RT) surface, our argument does not confirm they are indeed the RT surface. Our discussions are more qualitative. 
%%%%%%%%%%%%
%%%%%%%%%%%%
\subsubsection{Radial coordinate in gravity vs QFT}\label{sec:radial_coordinate}
\hspace{0.51cm}
%%%%%%%%%%%%
%%%%%%%%%%%%
Suppose there is an excitation localized in the bulk geometry. Then, we argue that, if we assume the standard AdS/CFT dictionary, the `radial coordinate' determined from $X_I$ indeed corresponds to the radial coordinate of the excitation on the gravity side. 

We do this by relating certain correlation functions to the radial coordinate. As a simple but instructive example, let us consider the Gaussian matrix model. For the ground state, we have
\begin{align}
\bra{\rm g.s.}
\hat{X}_{I,ij}
\ket{\rm g.s.}
=
0
\end{align}
and
\begin{align}
\bra{\rm g.s.}
\mathrm{Tr}\hat{X}_{I}^2
\ket{\rm g.s.}
\sim
N^2\, . 
\end{align}
For a low-energy wave packet (coherent state) around $Y_{I,ij}\neq 0$, we have
\begin{align}
\bra{Y,Q}
\hat{X}_{I,ij}
\ket{Y,Q}
=
Y_{I,ij}
\end{align}
and
\begin{align}
\bra{Y,Q}
\mathrm{Tr}\hat{X}_{I}^2
\ket{Y,Q}
=
\bra{\rm g.s.}
\mathrm{Tr}\hat{X}_{I}^2
\ket{\rm g.s.}
+
\mathrm{Tr}Y_I^2\, . 
\end{align}
If we take a generic excitation that is gauge-invariant by symmetrizing the coherent state as \eqref{eq:SU(N)-symmetrization}, the one-point function of $\hat{X}_{I,ij}$ becomes zero. Therefore, it is more convenient to use $\mathrm{Tr}\hat{X}_{I}^2$ as 

\begin{align}
\bra{Y,Q}
\mathrm{Tr}\hat{X}_{I}^2
\ket{Y,Q}
-
\bra{\rm g.s.}
\mathrm{Tr}\hat{X}_{I}^2
\ket{\rm g.s.}
=
\mathrm{Tr}Y_I^2\, . 
\end{align}
This relation holds even when $\ket{Y,Q}$ is replaced with its SU($N$)-symmetrized counterpart. 
Historically, people tried to use the expectation value of $\mathrm{Tr}\hat{X}_{I}^2$ to estimate the radial coordinate and got confused about the fact that this is of order $N^2$ already for the ground state. What we have seen above is we must subtract the vacuum contribution that contains only quantum fluctuations. The same strategy works for interacting theory as long as the back-reaction of small excitations to the ground state is negligible. Note that the use of $\mathrm{Tr}\hat{X}_{I}^2$ is not essential. Other functions of scalars could also work. 

We want to repeat a similar computation for QFT. For a product state of the form $\prod_{\vec{x}}(\int d\phi_{\vec{x}}f_{\vec{x}}(\phi_{\vec{x}})\ket{\phi_{\vec{x}}})$, we can obtain the information about the location of the excitations in the emergent space from the expectation value of $\mathrm{Tr}\hat{X}_I^2(\vec{x})$. This is because, roughly speaking, we can treat each spatial point as a different matrix model. The actual situation of interest can be more complicated because there is entanglement along the QFT space. Still, we expect at least a qualitative estimate of the location of the excitation from this quantity because of the local nature of the QFT space. 

\begin{figure}[htbp]
\begin{center}
\scalebox{0.25}{
\includegraphics{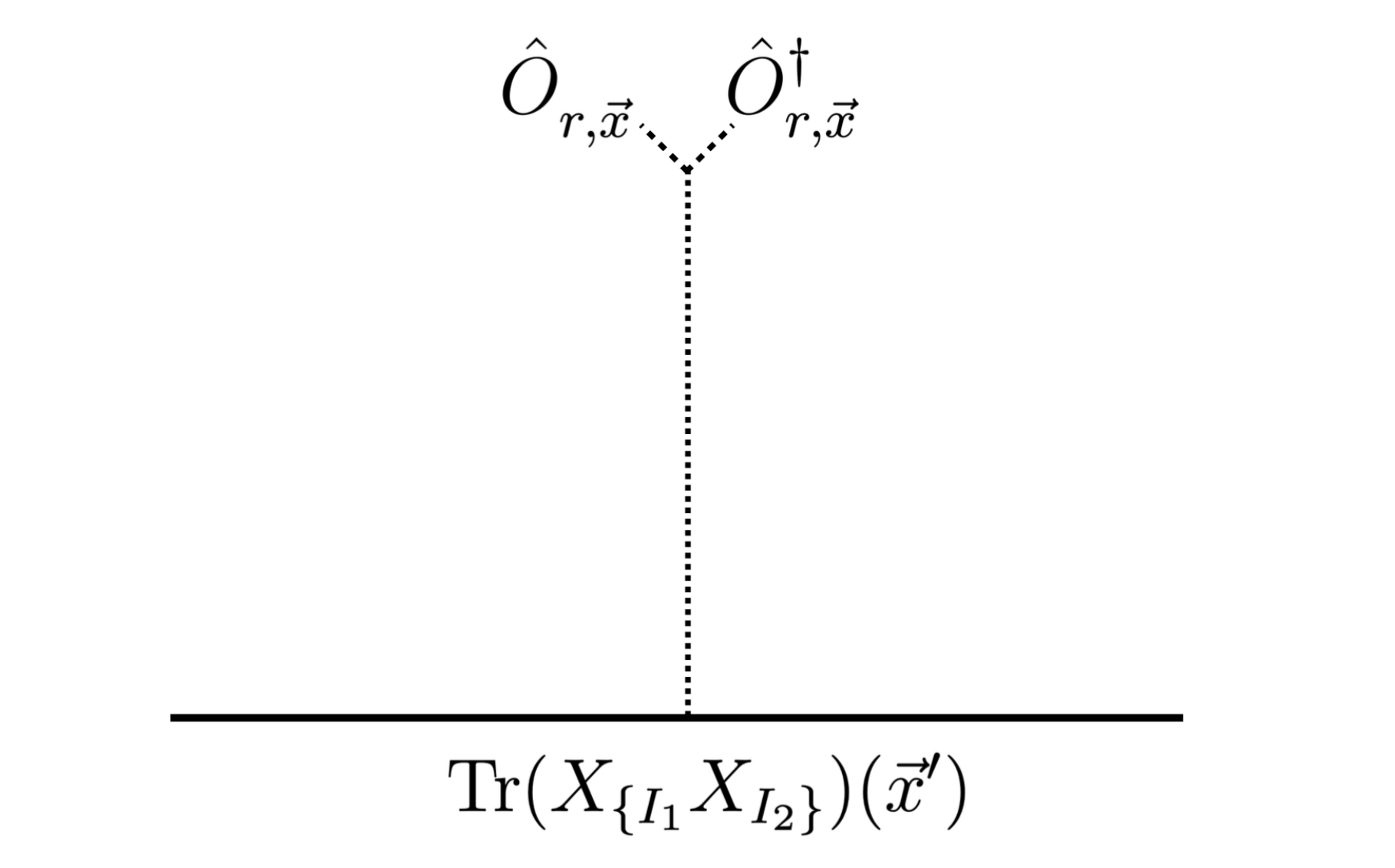}}
\end{center}
\caption{Schematic picture of connected three-point function in \eqref{eq:3-pt-fnc}. Because $\hat{O}$ and $\hat{O}^\dagger$ are at the same point in the bulk, we only need a three-point vertex and bulk-to-boundary two-point correlation function. 
}\label{3-pt-fnc}
\end{figure}

To simplify the computation by using supergravity, we focus on chiral primary operators. Let $\hat{O}_{r,\vec{x}}$ be a smeared operator that creates an excitation localized at a bulk point $(r,\vec{x})$. Note that we do not care exactly how such an operator is constructed: we only assume the existence of such an operator and compute correlation functions using the standard AdS/CFT dictionary. 
We consider a normalized state $\mathcal{N}^{-1/2}\hat{O}_{r,\vec{x}}\ket{\rm g.s.}$, where $\mathcal{N}=\bra{\rm g.s.}\hat{O}_{r,\vec{x}}^\dagger\hat{O}_{r,\vec{x}}\ket{\rm g.s.}$. 
We use the rank-2 chiral primary operator $\mathrm{Tr}(X_{\{I_1}X_{I_2\}})\equiv\mathrm{Tr}(X_{I_1}X_{I_2})-\frac{\delta_{I_1I_2}}{6}\sum_k\mathrm{Tr}(X_{k}X_{k})$ instead of $\mathrm{Tr}\hat{X}_{I}^2$, and take $\hat{O}$ the one that becomes the bulk-couterpart of the rank-$k_O$ chiral primary operator as $r\to\infty$. Deep inside the bulk, this operator can be complicated, but we do not need the explicit form below.
We should estimate\footnote{
The second term vanishes for chiral primaries.
}
\begin{align}
\frac{
\bra{\rm g.s.}\hat{O}^\dagger_{r,\vec{x}}
\mathrm{Tr}(X_{\{I_1}X_{I_2\}})(\vec{x}')
\hat{O}_{r,\vec{x}}\ket{\rm g.s.}
}{
\bra{\rm g.s.}\hat{O}^\dagger_{r,\vec{x}}\hat{O}_{r,\vec{x}}\ket{\rm g.s.}
}
-
\bra{\rm g.s.}\mathrm{Tr}(X_{\{I_1}X_{I_2\}})(\vec{x}')\ket{\rm g.s.}\, . 
\end{align}
Equivalently, we study the connected part of the 3-point correlator 
\begin{align}
\frac{
\bra{\rm g.s.}\hat{O}^\dagger_{r,\vec{x}}
\mathrm{Tr}(X_{\{I_1}X_{I_2\}})(\vec{x}')
\hat{O}_{r,\vec{x}}\ket{\rm g.s.}_{\rm conn}
}{
\bra{\rm g.s.}\hat{O}^\dagger_{r,\vec{x}}\hat{O}_{r,\vec{x}}\ket{\rm g.s.}
}\, . 
\label{eq:3-pt-fnc}
\end{align}
We can calculate it by using the Euclidean path integral at zero temperature, inserting the operators at Euclidean time zero. An important point is that $\mathrm{Tr}(X_{\{I_1}X_{I_2\}})$ is sitting at the boundary $(r'=\infty,\vec{x}')$ from the bulk point of view; see Fig.~\ref{3-pt-fnc}. 

To compute the three-point function, we insert a bulk three-point vertex that connects $\hat{O}$ and $\hat{O}^\dagger$ to $\mathrm{Tr}(X_{\{I_1}X_{I_2\}})$. To see the $r$-dependence, we can use the bulk-to-boundary two-point correlation function for $\mathrm{Tr}(X_{\{I_1}X_{I_2\}})$. Hence
\begin{align}
\frac{
\bra{\rm g.s.}\hat{O}^\dagger_{r,\vec{x}}
\mathrm{Tr}(X_{\{I_1}X_{I_2\}})(\vec{x}')
\hat{O}_{r,\vec{x}}\ket{\rm g.s.}_{\rm conn}
}{
\bra{\rm g.s.}\hat{O}^\dagger_{r,\vec{x}}\hat{O}_{r,\vec{x}}\ket{\rm g.s.}
}
\sim
r^2\cdot\left(\frac{1}{1+r^2(\vec{x}-\vec{x}')^2}\right)^{2}\, . 
\label{eq:3pt-scaling}
\end{align}
More generally, by replacing $\mathrm{Tr}(X_{\{I_1}X_{I_2\}})$ with the rank-$k$ chiral primary operator, we have  
\begin{align}
\frac{
\bra{\rm g.s.}\hat{O}^\dagger_{r,\vec{x}}
\mathrm{Tr}(X_{\{I_1}\cdots X_{I_k\}})(\vec{x}')
\hat{O}_{r,\vec{x}}\ket{\rm g.s.}_{\rm conn}
}{
\bra{\rm g.s.}\hat{O}^\dagger_{r,\vec{x}}\hat{O}_{r,\vec{x}}\ket{\rm g.s.}
}
\sim
r^k\cdot\left(\frac{1}{1+r^2(\vec{x}-\vec{x}')^2}\right)^{k}\, . 
\label{eq:3pt-scaling-2}
\end{align}
This is $r^k$ at $\vec{x}-\vec{x}'=0$ and decays quickly at $|\vec{x}-\vec{x}'|\gtrsim r^{-1}$. 
From this, we can see that the excitation sits at $\textrm{radial\ coordinate}\sim r$ in the emergent dimensions and localized at $|\vec{x}-\vec{x}'|\lesssim r^{-1}$ in the QFT space. Note that the same scaling can hold as long as $\hat{O}$ is a local operator in the bulk, not just for chiral primaries. 
If $\hat{O}$ is obtained by smearing the rank-$k_{O}$ chiral primary operator, a $k_{O}\times k_{O}$ block and $k_{O}$ eigenvalues can be excited; see \eqref{eq:excitation_block_size}. According to our proposal of the emergent space from scalars, these $k_{O}$ eigenvalues should scale as $r$, and hence, \eqref{eq:3pt-scaling} and \eqref{eq:3pt-scaling-2} should increase linearly with $k_{O}$. This is indeed the case~\cite{Lee:1998bxa}. 
See Appendix~\ref{sec:3pt-function} for details of the computations.  

Typically, a recipe for the bulk reconstruction is provided by assuming the AdS/CFT dictionary and reverse-engineering the CFT states from the `answer' on the gravity side. The argument above tells us where in the emergent dimensions from matrices such CFT operators are sitting. Therefore, our analysis provides a consistency check: if we assume the standard AdS/CFT dictionary, the `radial coordinate' determined from $X_I$ indeed corresponds to the radial coordinate of the excitation on the gravity side.  

%%%%%%%%%%%%
%%%%%%%%%%%%
\subsubsection{Size of boundary region vs radial coordinate}\label{sec:radial-coordinate-qualitative}
\hspace{0.51cm}
%%%%%%%%%%%%
%%%%%%%%%%%%
Widely believed pictures of bulk reconstruction suggest that the size of the region on the boundary where the bulk wave packet has support is related to the radial coordinate of AdS. 
The argument in Sec.~\ref{sec:radial_coordinate} explains this connection assuming that the previous results based on computations on the gravity side are correct.  
In this section, we would like to understand this relation without relying on the gravity dual.
We consider a QFT wave packet, which should be understood as a linear combination of bulk wave packets, and show that the relation between the size of the wave packet in the QFT space and the radial coordinate defined by scalars is qualitatively consistent with the holographic dictionary. 
We emphasize that we consider particularly simple QFT wave packets so that the computations are feasible without relying on gravity dual. Excitations well localized in the bulk require more complicated form.
%%%%%%%%%%%%%%%%%
%%%%%%%%%%%%%%%%%
\subsubsection*{Intuition from lattice regularization}
%%%%%%%%%%%%%%%%%
%%%%%%%%%%%%%%%%% 
Let us use lattice regularization on a spatial lattice with lattice spacing $a$. An excitation localized on only one point on the lattice is analogous to a wave packet of size $a$ in the continuum theory. 
With $d$ spatial dimensions, we can consider the square lattice with equal lattice spacing $a$ in all $d$ directions.
The lattice Hamiltonian is defined by
\begin{align}\label{eq:2d-lattice-Hamiltonian}
    \hat{H}_{\rm lattice}
    =
    a\hat{H}
    =
    \sum_{\vec{n}}\left(
    \frac{1}{2}
    \hat{\pi}_{{\rm lattice},\vec{n}}^2
    +
    \frac{1}{2}
    \sum_{\mu=1}^d
    \left(
    \hat{\phi}_{{\rm lattice},\vec{n}+\hat{\mu}}-\hat{\phi}_{{\rm lattice},\vec{n}}
    \right)^2
    +
    V(\hat{\phi}_{{\rm lattice},\vec{n}})
    \right) \, . 
\end{align}
$\hat{\mu}$ is the unit vector along the $\mu$-th dimension of the spatial lattice ($\mu=1,\cdots,d$).
We have the canonical commutation relation 
\begin{align}
[\hat{\phi}_{{\rm lattice},\vec{n}},
\hat{\pi}_{{\rm lattice},\vec{n}'}]
=
i\delta_{\vec{n}\vec{n}'}\, .  
\end{align}
Fields $\hat{\phi}_{\rm lattice}$ and $\hat{\pi}_{\rm lattice}$ are dimensionless. 
To take the continuum limit, dimensions should be restored as 
\begin{align}
\hat{\phi}_{\rm lattice}
=
a^{(d-1)/2}\hat{\phi}\, , 
\qquad
\hat{\pi}_{\rm lattice}
=
a^{(d+1)/2}\hat{\pi}\, . 
\end{align}
Roughly speaking, $\phi$ in the lattice theory is average of $\phi(\vec{x})$ in the continuum theory over the unit lattice, $a^{-d}\int_{|\vec{x}'-\vec{x}|<a} d^d\vec{x}'\phi(\vec{x}')$. 
Indeed, we can see that this $a$ corresponds to the width of the Gaussian wave packet $\sigma$ in \eqref{eq:Gaussian-WP-commutator}. 

The lattice Hamiltonian $\hat{H}_{\rm lattice}$ can depend on lattice spacing $a$ via couplings in $V(\hat{\phi}_{{\rm lattice},\vec{n}})$, i.e., dimensionful couplings such as mass $m^2$ is made dimensionless by multiplying appropriate power of $a$, e.g., $m^2a^2$. In addition, there can be a running of couplings associated with renormalization. 

\begin{figure}[htbp]
\begin{center}
\scalebox{0.5}{
\includegraphics{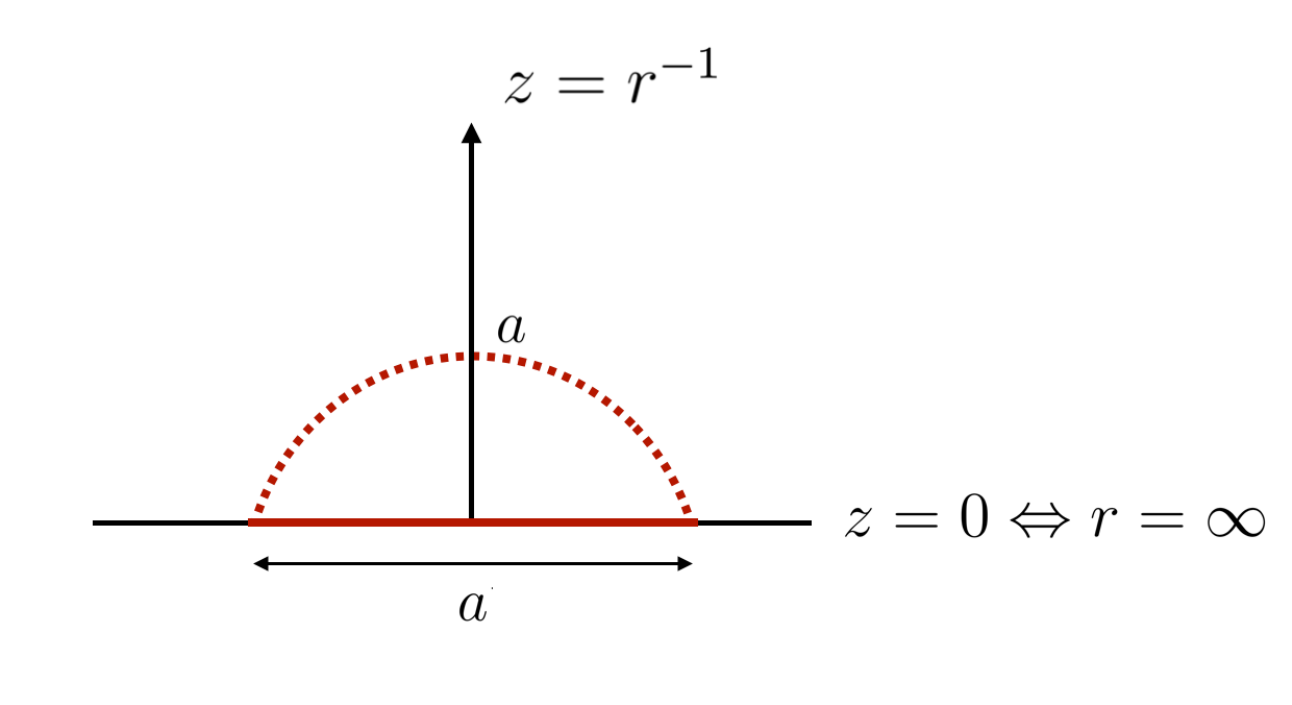}}
\end{center}
\caption{Operators that has a support in a region of size $a^3$ in the QFT space can create bulk wave packets at $r\gtrsim a^{-1}$ (equivalently, $z\lesssim a$). Note that $r=\infty$ is the boundary. 
}\label{wave-packet-QFT}
\end{figure}

Suppose we could ignore the explicit $a$-dependence in $V(\hat{\phi}_{{\rm lattice},\vec{n}})$, e.g., in a conformal theory such as 4d $\mathcal{N}=4$ SYM.\footnote{Strictly speaking, we need to include dynamical gauge fields and fermions to the lattice action. Explicit $a$-dependence can be avoided by using fine-tuning-free formulations of 4d SYM~\cite{Hanada:2010kt,Hanada:2010gs}.
} Then, the location of the center of lowest-energy wave packet made of a polynomial of $\hat{\phi}_{{\rm lattice},\vec{n}}$ has to be of order $1$ in terms of $\hat{\phi}_{{\rm lattice},\vec{n}}$ and $a^{-(d-1)/2}$ (which is $a^{-1}$ for $d=3$) in terms of $\hat{\phi}$; see Fig.~\ref{wave-packet-QFT}. 
This is the same as the scaling we observed in Sec.~\ref{sec:radial_coordinate}, by identifying $a$ with $r^{-1}$. 
%%%%%%%%%%%%%%%%%
%%%%%%%%%%%%%%%%%
\subsubsection*{Gaussian wave packet in the weak-coupling limit}
%%%%%%%%%%%%%%%%%
%%%%%%%%%%%%%%%%%
Let us see how the logic above works for the simplest explicit example: the Gaussian-smeared operators \eqref{eq:Gaussian_smeared_operators} acting on the ground state of massless free scalar theory on $\mathbb{R}^3$. The Hamiltonian is 
\begin{align}
\hat{H}_{\rm free}
&=
\frac{1}{2}\int d^3x
\left(
\hat{\pi}_{\vec{x}}^2
+
\sum_{j=1}^3
(\partial_j\hat{\phi}_{\vec{x}})^2
\right)\, .
\end{align}
In the momentum basis, the Hamiltonian is a collection of harmonic oscillators:
\begin{align}
\hat{H}_{\rm free}
&=
 \frac{1}{2}\int d^3\vec{k}
\left(
\hat{\tilde{\pi}}_{\vec{k}}
\hat{\tilde{\pi}}_{-\vec{k}}
+
\vec{k}^2\cdot
\hat{\tilde{\phi}}_{\vec{k}}
\hat{\tilde{\phi}}_{-\vec{k}}
\right)
=
 \frac{1}{2}\int d^3\vec{k}
 \left(
\hat{\tilde{a}}^\dagger_{\vec{k}}
\hat{\tilde{a}}_{\vec{k}}
 +
 \frac{1}{2}
 \right)\omega_{\vec{k}}\, .
\end{align}
Here, 
\begin{align}
\phi_{\vec{x}}
=
\frac{1}{(2\pi)^{3/2}}\int d^3\vec{k} e^{ikx}\hat{\tilde{\phi}}_{\vec{k}}\, , 
\qquad
\hat{\tilde{\phi}}_{\vec{k}}
=
\frac{1}{(2\pi)^{3/2}}\int d^3\vec{x} e^{-ikx}\phi_{\vec{x}}\, , 
\nonumber\\
\pi_{\vec{x}}
=
\frac{1}{(2\pi)^{3/2}}\int d^3\vec{k} e^{ikx}\hat{\tilde{\pi}}_{\vec{k}}\, , 
\qquad
\hat{\tilde{\pi}}_{\vec{k}}
=
\frac{1}{(2\pi)^{3/2}}\int d^3\vec{x} e^{-ikx}\pi_{\vec{x}}\, , 
\end{align}
\begin{align}
\hat{\tilde{\phi}}_{\vec{k}}^\dagger
=
\hat{\tilde{\phi}}_{-\vec{k}}\, , 
\qquad
\hat{\tilde{\pi}}_{\vec{k}}^\dagger
=
\hat{\tilde{\pi}}_{-\vec{k}}\, . 
\end{align}
Note that the momentum conjugate to $\hat{\phi}_{\vec{k}}$ is $(\hat{\pi}_{\vec{k}})^\dagger=\hat{\pi}_{-\vec{k}}$, and the canonical commutation relation is 
\begin{align}
[\hat{\tilde{\phi}}_{\vec{k}},\hat{\tilde{\pi}}_{-\vec{k}'}]
=
i\delta^3(\vec{k}-\vec{k}')\, . 
\end{align}
The annihilation operator $\hat{\tilde{a}}_{\vec{k}}$ and creation operator $\hat{\tilde{a}}^\dagger_{\vec{k}}$ are defined as
\begin{align}
\hat{\tilde{a}}_{\vec{k}}
=
\frac{
\sqrt{\omega_{\vec{k}}}\hat{\tilde{\phi}}_{\vec{k}}
+
\frac{i}{\sqrt{\omega_{\vec{k}}}}\hat{\tilde{\pi}}_{\vec{k}}
}{\sqrt{2}}\, , 
\qquad
\hat{\tilde{a}}_{\vec{k}}^\dagger
=
\frac{
\sqrt{\omega_{\vec{k}}}\hat{\tilde{\phi}}_{-\vec{k}}
-
\frac{i}{\sqrt{\omega_{\vec{k}}}}\hat{\tilde{\pi}}_{-\vec{k}}
}{\sqrt{2}}\, , 
\end{align}
where $\omega_{\vec{k}}=\sqrt{\vec{k}^2}$. 
They satisfy
\begin{align}
[\hat{\tilde{a}}_{\vec{k}},\hat{\tilde{a}}_{\vec{k}'}^\dagger]
=
\delta^3(\vec{k}-\vec{k}')\, . 
\end{align}
The ground state is the Fock vacuum $\ket{0}$ that is annihilated by all annihilation operators:
\begin{align}
\hat{\tilde{a}}_{\vec{k}}\ket{0}=0\, . 
\end{align}

In the momentum basis, the Gaussian-smeared operators \eqref{eq:Gaussian_smeared_operators} are written as
\begin{align}
\hat{\Phi}_{\vec{x}_0,\vec{p}_0}
&=
\frac{1}{(2\pi)^{3/2}}
\int d^3\vec{k}
e^{-\frac{\sigma^2}{2}(\vec{k}-\vec{p}_0)^2+i\vec{k}\cdot\vec{x}_0}\hat{\tilde{\phi}}_{\vec{k}}
\nonumber\\
&=
\frac{1}{(2\pi)^{3/2}}
\int d^3\vec{k}
e^{-\frac{\sigma^2}{2}(\vec{k}-\vec{p}_0)^2+i\vec{k}\cdot\vec{x}_0}
\cdot
\frac{\hat{\tilde{a}}_{\vec{k}}+\hat{\tilde{a}}^\dagger_{-\vec{k}}}{\sqrt{2\omega_{\vec{k}}}}\, , 
\end{align}
etc. 
Acting $\hat{\Phi}_{\vec{x}_0,\vec{p}_0}$ on the Fock vacuum, we obtain 
\begin{align}
\ket{\Psi}
\equiv
\hat{\Phi}_{\vec{x}_0,\vec{p}_0}\ket{0}
=
\frac{1}{(2\pi)^{3/2}}
\int d^3\vec{k}
e^{-\frac{\sigma^2}{2}(\vec{k}-\vec{p}_0)^2+i\vec{k}\cdot\vec{x}_0}
\cdot
\frac{\hat{\tilde{a}}^\dagger_{-\vec{k}}}{\sqrt{2\omega_{\vec{k}}}}
\ket{0}\, . 
\end{align}
Following the idea presented in Sec.~\ref{sec:radial_coordinate}, let us compute 
\begin{align}
\frac{\bra{\Psi}\hat{\phi}_{\vec{x}}^2\ket{\Psi}}{\bra{\Psi}\ket{\Psi}}
-
\bra{0}\hat{\phi}_{\vec{x}}^2\ket{0}\, . 
\label{eq:excitation_free_theory}
\end{align}
The ground-state contribution is 
\begin{align}
\bra{0}\hat{\phi}_{\vec{x}}^2\ket{0}
&=
\frac{1}{(2\pi)^3}\int d^3\vec{k}\int d^3\vec{k}' e^{i(\vec{k}+\vec{k}')\cdot\vec{x}}
\bra{0}\hat{\tilde{\phi}}_{\vec{k}}\hat{\tilde{\phi}}_{\vec{k}'}\ket{0}
\nonumber\\
&=
\frac{1}{(2\pi)^3}\int d^3\vec{k}\int d^3\vec{k}' e^{i(\vec{k}+\vec{k}')\cdot\vec{x}}
\bra{0}
\frac{\hat{\tilde{a}}_{\vec{k}}}{\sqrt{2\omega_{\vec{k}}}}
\cdot
\frac{\hat{\tilde{a}}_{-\vec{k}'}^\dagger}{\sqrt{2\omega_{\vec{k}'}}}
\ket{0}
\nonumber\\
&=
\frac{1}{(2\pi)^3}\int d^3\vec{k}\int d^3\vec{k}' e^{i(\vec{k}+\vec{k}')\cdot\vec{x}}
\cdot
\frac{
\delta^3(\vec{k}+\vec{k}')
}{
2\sqrt{\omega_{\vec{k}}\omega_{\vec{k}'}}
}
\nonumber\\
&=
\frac{1}{(2\pi)^3}\int\frac{d^3\vec{k}}{2k}\, , 
\end{align}
which is ultraviolet divergent. This term cancels with the counterpart in the first term of \eqref{eq:excitation_free_theory}.
The normalization factor $\bra{\Psi}\ket{\Psi}$ is 
\begin{align}
\bra{\Psi}\ket{\Psi}
&=
\frac{1}{(2\pi)^3}
\int d^3\vec{k}
\frac{e^{-\sigma^2(\vec{k}-\vec{p}_0)^2}}{2\omega_k}\, . 
\end{align}
For $\vec{p}_0=\vec{0}$, we have
\begin{align}
\bra{\Psi}\ket{\Psi}_{\vec{p}_0=\vec{0}}
&=
\frac{1}{(2\pi)^3}
\int_0^\infty 4\pi k^2dk
\frac{e^{-\sigma^2k^2}}{2k}
=
\frac{1}{8\pi^2\sigma^2}\, . 
\end{align}
Finally, 
\begin{align}
\bra{\Psi}\hat{\phi}_{\vec{x}}^2\ket{\Psi}
%%%
&=
\frac{1}{(2\pi)^6}
\int d^3\vec{k}
e^{-\frac{\sigma^2}{2}(\vec{k}-\vec{p}_0)^2+i\vec{k}\cdot\vec{x}_0}
\int d^3\vec{k}'
e^{-\frac{\sigma^2}{2}(\vec{k}'-\vec{p}_0)^2-i\vec{k}'\cdot\vec{x}_0}
\nonumber\\
&\qquad
\int d^3\vec{l}\int d^3\vec{l}' e^{i(\vec{l}+\vec{l}')\cdot\vec{x}}
\bra{0}
\frac{\hat{\tilde{a}}_{-\vec{k}'}}{\sqrt{2\omega_{\vec{k}'}}}
\cdot
\frac{\hat{\tilde{a}}_{\vec{l}}+\hat{\tilde{a}}_{-\vec{l}}^\dagger
}{\sqrt{2\omega_{\vec{l}}}}
\cdot
\frac{\hat{\tilde{a}}_{\vec{l}'}+\hat{\tilde{a}}_{-\vec{l}'}^\dagger
}{\sqrt{2\omega_{\vec{l}'}}}
\cdot
\frac{\hat{\tilde{a}}^\dagger_{-\vec{k}}}{\sqrt{2\omega_{\vec{k}}}}
\ket{0}
\nonumber\\
%%%
&=
\frac{1}{(2\pi)^6}
\int d^3\vec{k}
e^{-\frac{\sigma^2}{2}(\vec{k}-\vec{p}_0)^2+i\vec{k}\cdot\vec{x}_0}
\int d^3\vec{k}'
e^{-\frac{\sigma^2}{2}(\vec{k}'-\vec{p}_0)^2-i\vec{k}'\cdot\vec{x}_0}
\nonumber\\
&\qquad
\int d^3\vec{l}\int d^3\vec{l}' \frac{e^{i(\vec{l}+\vec{l}')\cdot\vec{x}}}{4\sqrt{\omega_{\vec{k}}\omega_{\vec{k}'}\omega_{\vec{l}}\omega_{\vec{l}'}}}
\nonumber\\
&\qquad
\times
\left\{
\delta^3(\vec{l}+\vec{l}')\delta^3(\vec{k}-\vec{k}')
+
\delta^3(\vec{k}+\vec{l}')\delta^3(-\vec{k}'+\vec{l})
+
\delta^3(\vec{k}+\vec{l})\delta^3(-\vec{k}'+\vec{l}')
\right\}\, . 
\end{align}
From $\delta^3(\vec{l}+\vec{l}')\delta^3(\vec{k}-\vec{k}')$, we obtain
\begin{align}
\frac{1}{(2\pi)^6}
\int d^3\vec{k}
e^{-\sigma^2(\vec{k}-\vec{p}_0)^2}
\int d^3\vec{l}\frac{1}{4\omega_{\vec{k}}\omega_{\vec{l}}}
&=
\bra{\Psi}\ket{\Psi}
\cdot
\frac{1}{(2\pi)^3}\int\frac{d^3\vec{l}}{2l}\, , 
\end{align}
which cancels with $\bra{0}\hat{\phi}_{\vec{x}}^2\ket{0}$. 
From the other terms, we obtain
\begin{align}
\frac{1}{(2\pi)^6}
\int d^3\vec{k}
e^{-\frac{\sigma^2}{2}(\vec{k}-\vec{p}_0)^2-i\vec{k}\cdot(\vec{x}-\vec{x}_0)}
\int d^3\vec{k}'
e^{-\frac{\sigma^2}{2}(\vec{k}'-\vec{p}_0)^2+i\vec{k}'\cdot(\vec{x}-\vec{x}_0)}
\cdot
\frac{1}{4\omega_{\vec{k}}\omega_{\vec{k}'}}
\end{align}
This takes maximum at $\vec{x}=\vec{x}_0$. Furthermore, if we assume $\vec{p}_0=\vec{0}$ for simplicity, we obtain
\begin{align}
\frac{1}{(2\pi)^6}
\left(
\int d^3\vec{k}
\frac{e^{-\frac{\sigma^2}{2}\vec{k}^2-i\vec{k}\cdot(\vec{x}-\vec{x}_0)}}{2\omega_{\vec{k}}}
\right)^2
=
\frac{1}{(2\pi)^6}
\times
\left(
\frac{2\pi}{\sigma^2}
e^{-\frac{(\vec{x}-\vec{x}_0)}{2\sigma^2}}
\right)^2
%=
%\frac{e^{-\frac{(\vec{x}-\vec{x}_0)}{2\sigma^2}}}{(2\pi)^7\sigma^{10}}
=
\frac{e^{-\frac{(\vec{x}-\vec{x}_0)}{\sigma^2}}}{4\pi^2\sigma^2}
\cdot
\bra{\Psi}\ket{\Psi}\, . 
\end{align}
Therefore, for $\vec{p}_0=\vec{0}$, \eqref{eq:excitation_free_theory} becomes
\begin{align}
\frac{\bra{\Psi}\hat{\phi}_{\vec{x}}^2\ket{\Psi}}{\bra{\Psi}\ket{\Psi}}
-
\bra{0}\hat{\phi}_{\vec{x}}^2\ket{0}
=
\frac{e^{-\frac{(\vec{x}-\vec{x}_0)}{\sigma^2}}}{4\pi^2\sigma^2}\, . 
\end{align} 
Identifying $\sigma$, $a$ and $r^{-1}$, we see qualitatively the same scaling as before. 
%%%%%%%%%%%%%%%%%
%%%%%%%%%%%%%%%%%
\subsubsection{Comment on ($2+1$)-dimensional theory}\label{sec:ABJM}
%%%%%%%%%%%%%%%%%
%%%%%%%%%%%%%%%%%
Repeating the arguments in Sec.~\ref{sec:radial-coordinate-qualitative} for $d=2$, we obtain `radial coordinate' $a^{-(d-1)/2}=a^{-1/2}$ for a QFT wave packet of size $a$. This scaling is natural in the following sense. 

In the ABJM theory~\cite{Aharony:2008ug} with $\mathrm{U}(N)\times\mathrm{U}(N)$ gauge group, there are four bi-fundamental scalars $C_{\alpha=1,\cdots,4}$. 
The potential term of the scalars is proportional to
\begin{align}
-\frac{1}{6}\mathrm{Tr}(C_\alpha\bar{C}^\alpha)^3
-\frac{1}{6}\mathrm{Tr}
(\bar{C}^\alpha C_\alpha)^3
-
\frac{2}{3}\mathrm{Tr}(C_\alpha\bar{C}^\gamma C_\beta\bar{C}^\alpha C_\gamma\bar{C}^\beta)
+
\mathrm{Tr}(C_\alpha\bar{C}^\alpha C_\beta\bar{C}^\gamma C_\gamma\bar{C}^\beta)\, . 
\end{align}
Note that a product $C_\alpha\bar{C}^\beta$ transforms in the adjoint representation of the first U($N$), analogous to scalars in 4d SYM. 
The scaling is $C_\alpha\sim a^{-1/2}$ means $C_\alpha\bar{C}^\beta\sim a^{-1}$, which leads to natural energy scale. For example, assuming $C_1$ is diagonal, i.e., $C_1=D\equiv\mathrm{diag}(d_1,\cdots,d_N)$, the scalar potential is proportional to 
\begin{align}
\sum_{\alpha=2}^4\mathrm{Tr}\left([D\bar{D},C_\alpha][D\bar{D},\bar{C}^\alpha]\right)\,  
\end{align}
up to terms of higher order in $C_{\alpha=2,3,4}$. Therefore, $D\sim a^{-1/2}$ means the mass of off-diagonal entries of order $a^{-1}$, as in the case of 4d SYM.
%%%%%%%%%%%%%%%%%
%%%%%%%%%%%%%%%%%
\subsubsection{Bulk wave packet and quantum entanglement}\label{sec:geometry_from_entanglement}
\begin{figure}[htbp]
\begin{center}
\scalebox{0.2}{
\includegraphics{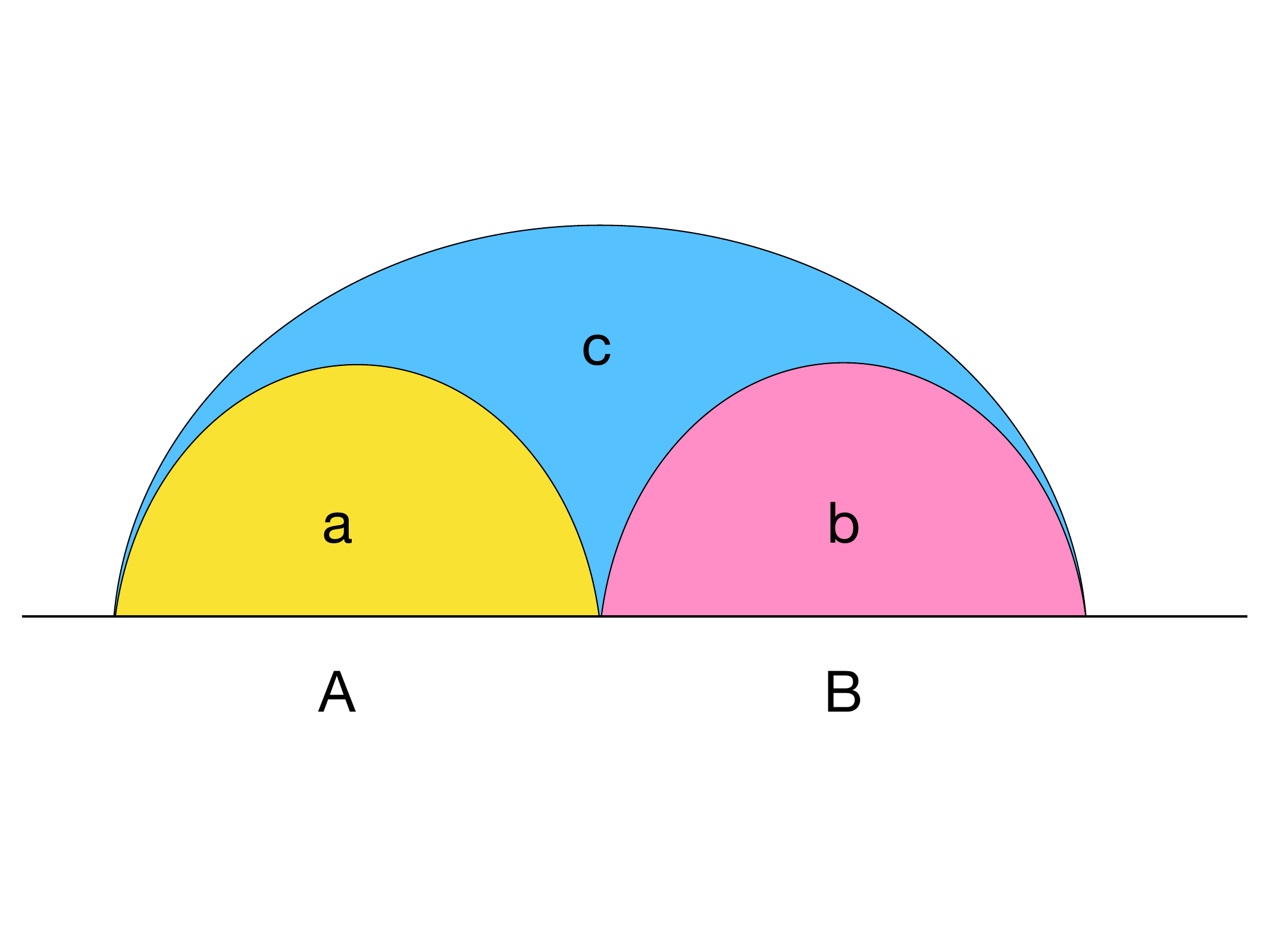}}
\scalebox{0.2}{
\includegraphics{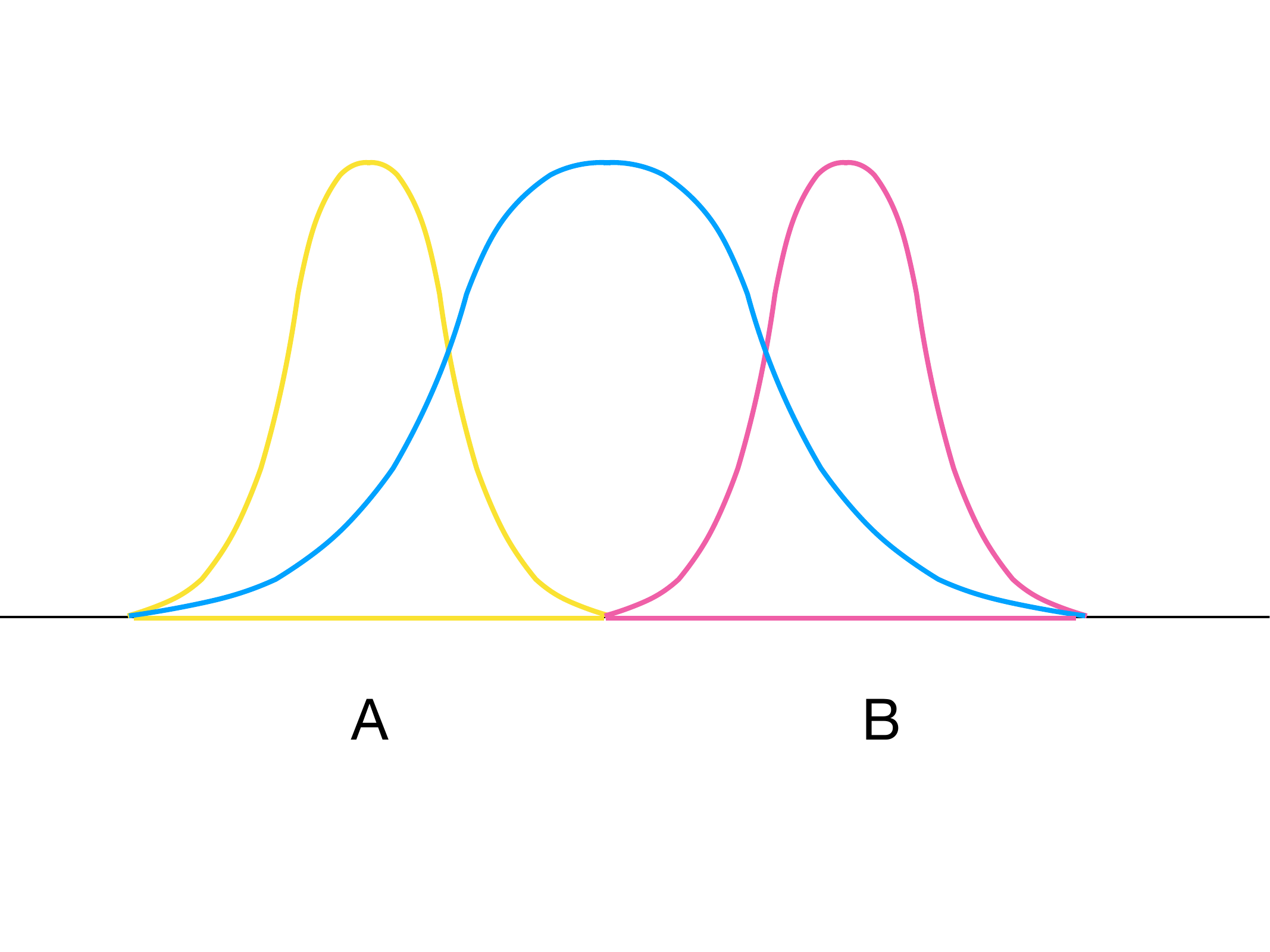}}
\end{center}
\caption{[Left] bulk region c is generated due to the entanglement between two boundary regions A and B.
[Right] A cartoon picture of smearing functions. The vertical axis is $f(\vec{x})$ in \eqref{eq:A-B-entanglement}. To create a bulk wave packet in a, b, or c, $f(\vec{x})$ should have support in A, B, or A$\cup$B like the line with the corresponding color. The important point is that such a figure could be drawn without referring to gravity dual, identifying the `radial coordinate' with the value of scalar field, or more specifically, the location of the bulk wave packet.    
}\label{entanglement}
\end{figure}

\begin{figure}[htbp]
\begin{center}
\scalebox{0.2}{
\includegraphics{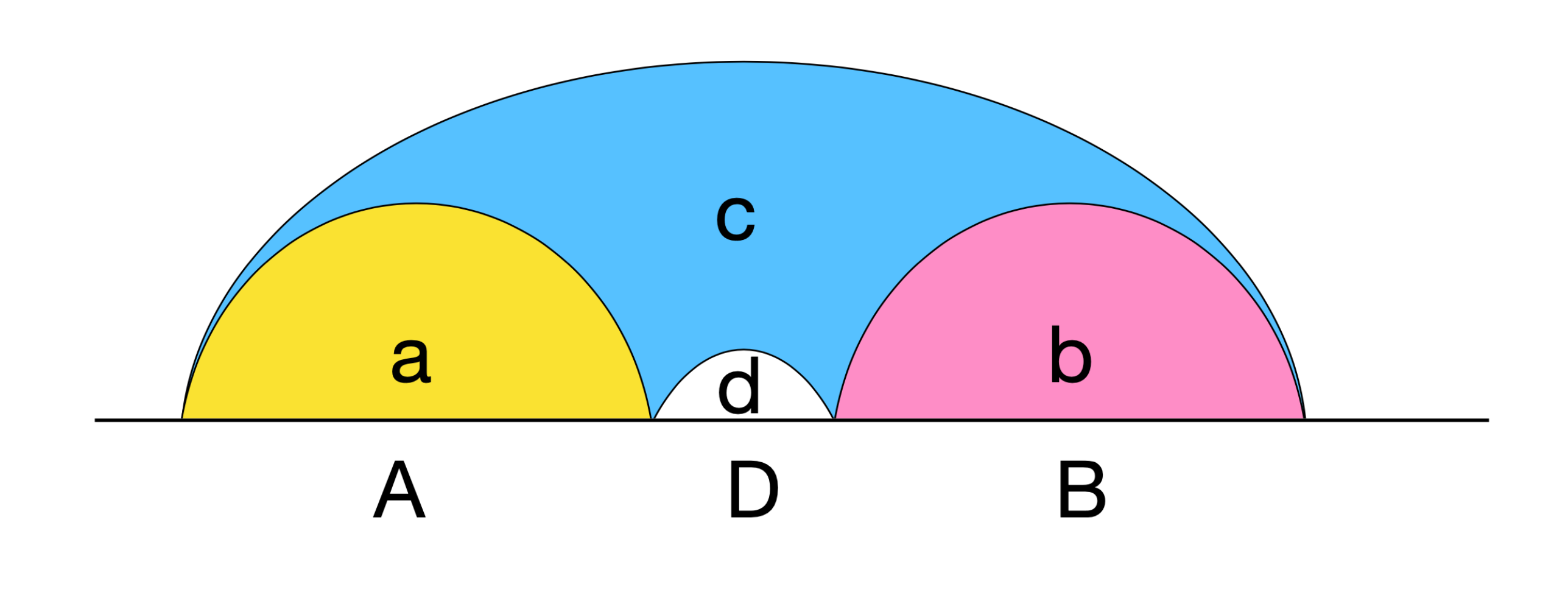}}
\end{center}
\caption{Region c is generated due to the entanglement between two disjoint regions $A$ and $B$  in the boundary.
Region c is made of low-frequency modes ($f(\vec{x})$ with large support). 
Region d requires high-frequency modes ($f(\vec{x})$ with small support) and hence cannot be generated. 
}\label{entanglement_disjoint_regions}
\end{figure}

%%%%%%%%%%%%%%%%%
%%%%%%%%%%%%%%%%%
Let us consider the union of two regions on the boundary A and B, as shown in the left panel of Fig.~\ref{entanglement}. We can also consider a union of two disconnected regions of the boundary, as shown in Fig.~\ref{entanglement_disjoint_regions}. Either way, region c is analogous to a traversable wormhole as we will discuss in Sec.~\ref{sec:traversable-wormhole}. Below, we study the role of the entanglement of boundary regions in the creation of bulk local excitations. The entanglement discussed here is the $O(N^0)$ corrections on top of the $O(N^2)$ entanglement of the ground state. The role of the ground-state contribution is explained in Sec.~\ref{sec:traversable-wormhole}. 

In these figures, the vertical direction is the emergent space from scalars. A wave packet that has support in one of the boundary regions A or B can describe only an excitation in bulk regions a or b. In the right panel, we showed a schematic picture of such smearing functions that can create wave packets in the emergent bulk geometry. However, a wave packet that has support in the union of two boundary regions can describe an excitation in a bulk region c as well; see Fig.~\ref{entanglement}. Such a wave packet is created entangling the excitations in region A and region B. To understand this intuitively, imagine a pure state made of a QFT wave packet obtained by using a smearing function depicted by the blue line in the right panel of Fig.~\ref{entanglement}. 
We can write it trivially as 
\begin{align}
\left(\int_{{\rm A}\cup{\rm B}} d^3\vec{x}f(\vec{x})\hat{\phi}_{\vec{x}}\right)\ket{\rm g.s.}
=
\left(\int_{\rm A} d^3\vec{x}f(\vec{x})\hat{\phi}_{\vec{x}}\right)\ket{\rm g.s.}
+
\left(\int_{\rm B} d^3\vec{x}f(\vec{x})\hat{\phi}_{\vec{x}}\right)\ket{\rm g.s.}\, . 
\label{eq:A-B-entanglement}
\end{align}
(Here, $f(\vec{x})$ is a smearing function. We could use a generic composite operator instead of $\hat{\phi}_{\vec{x}}$.)
The excitation is either in boundary region A or boundary region B; this is an entangled state. 
As essentially the same case, let us construct an entangled state from one cat. Suppose that a cat wants to go into a box and there are two boxes, $A$ and $B$. Then, 
\begin{align}
\frac{\ket{\rm cat}_{\rm A}\otimes\ket{\rm no\ cat}_{\rm B}+\ket{\rm no\ cat}_{\rm A}\otimes\ket{\rm cat}_{\rm B}}{\sqrt{2}}
\label{eq:1-cat-entanglement}
\end{align}
is an entangled state made of only one cat. The similarity between this state and the right-hand side of \eqref{eq:A-B-entanglement} must be obvious. It would be instructive to see the difference between such an entangled state and more common entangled states made of two cats, e.g., 
\begin{align}
\frac{\ket{\rm white\ cat}_{\rm A}\otimes\ket{\rm black\ cat}_{\rm B}+\ket{\rm black\ cat}_{\rm A}\otimes\ket{\rm white\ cat}_{\rm B}}{\sqrt{2}}\, . 
\end{align}

The meaning of the entanglement of the form \eqref{eq:A-B-entanglement}, \eqref{eq:1-cat-entanglement} becomes clearer after seeing the similarity to the traverse wormhole, which is discussed in Sec.~\ref{sec:traversable-wormhole}. 

%%%%%%%%%%%%
%%%%%%%%%%%%
\subsubsection{Traversable wormhole and bulk wave packet}\label{sec:traversable-wormhole}
\hspace{0.51cm}
%%%%%%%%%%%%
%%%%%%%%%%%%
The duality between a two-sided black hole and thermofield double state~\cite{Maldacena:2001kr} is arguably the clearest example of the emergence of spacetime (in this case, wormhole) from entanglement~\cite{VanRaamsdonk:2010pw}. We suggest that this case can also be understood in terms of bulk wave packets, in the same way as in Sec.~\ref{sec:geometry_from_entanglement}.  

Let us consider a traversable wormhole. A well-known example is generated by the coupled SYK model~\cite{Maldacena:2018lmt}. (Different from a famous work that related traversable wormhole and quantum teleportation~\cite{Gao:2016bin}, we discuss an eternal traversable wormhole that is static and dual to the ground state.) To make the point even simpler, let us consider the simplest toy model: the coupled harmonic oscillator~\cite{Srednicki:1993im,Alet:2020ehp}. 
The Hamiltonian is
\begin{align}
\hat{H}
=
\frac{1}{2}\hat{p}_{\rm A}^2
+
\frac{1}{2}\hat{x}_{\rm A}^2
+
\frac{1}{2}\hat{p}_{\rm B}^2
+
\frac{1}{2}\hat{x}_{\rm B}^2
-
c\hat{x}_{\rm A}\hat{x}_{\rm B}\, , 
\end{align}
where $c$ is a small coupling.
By using new variables $\hat{X}_\pm=\frac{\hat{x}_{\rm A}\pm\hat{x}_{\rm B}}{\sqrt{2}}$ and $\hat{P}_\pm=\frac{\hat{p}_{\rm A}\pm\hat{p}_{\rm B}}{\sqrt{2}}$, we can write the Hamiltonian as a sum of two harmonic oscillators whose frequencies are shifted from the original value:
\begin{align}
\hat{H}
=
\frac{1}{2}\hat{P}_+^2
+
\frac{\omega_+^2}{2}\hat{X}_+^2
+
\frac{1}{2}\hat{P}_-^2
+
\frac{\omega_-^2}{2}\hat{X}_-^2\, , 
\end{align}
where
\begin{align}
\omega_\pm^2=1\mp c\, . 
\end{align}
The ground state is the Fock vacuum in terms of these new harmonic oscillators $\hat{X}_\pm$. 
In terms of the original oscillators $\hat{x}_{\rm A}$ and $\hat{x}_{\rm B}$, a thermofield double state is realized; see Refs.~\cite{Srednicki:1993im,Alet:2020ehp} for details.\footnote{
Strictly speaking, we need to use slightly different oscillators whose frequencies are slightly different from 1 to make the thermofield-double structure manifest~\cite{Alet:2020ehp}. This fact does not affect the following discussions. 
} The same holds for the coupled SYK model and the ground state is interpreted as two copies of AdS connected by traversable wormhole~\cite{Maldacena:2018lmt}.

Creation and annihilation operators $\hat{A}^\dagger_\pm$, $\hat{A}_\pm$ in the $\pm$-basis and those in the ${\rm A,B}$-basis $\hat{a}^\dagger_{\rm A,B}$, $\hat{a}_{\rm A,B}$ are related as follows:
\begin{align}
\hat{a}_{\rm A}^\dagger
=
\frac{\hat{x}_{\rm A}-i\hat{p}_{\rm A}}{\sqrt{2}}\, , 
\qquad
\hat{a}_{\rm B}^\dagger
=
\frac{\hat{x}_{\rm B}-i\hat{p}_{\rm B}}{\sqrt{2}}\, , 
\end{align}

\begin{align}
\hat{X}_\pm
=
\frac{\hat{a}_{\rm A}+\hat{a}_{\rm A}^\dagger\pm\hat{a}_{\rm B}\pm\hat{a}_{\rm B}^\dagger}{2}\, , 
\qquad
\hat{P}_\pm
=
\frac{\hat{a}_{\rm A}-\hat{a}_{\rm A}^\dagger\pm\hat{a}_{\rm B}\mp\hat{a}_{\rm B}^\dagger}{2i}\, , 
\end{align}

\begin{align}
\hat{A}_\pm^\dagger
=
\frac{\sqrt{\omega_\pm}\hat{X}_\pm-i\frac{\hat{P}_\pm}{\sqrt{\omega_\pm}}}{\sqrt{2}}
\simeq
\hat{a}^\dagger_\pm
\mp
\frac{c\hat{a}_\pm}{4}\, , 
\end{align}
where
\begin{align}
\hat{a}_\pm
\equiv
\frac{\hat{a}_{\rm A}
\pm
\hat{a}_{\rm B}}{\sqrt{2}}\, , 
\qquad
\hat{a}^\dagger_\pm
\equiv
\frac{\hat{a}_{\rm A}^\dagger
\pm
\hat{a}_{\rm B}^\dagger}{\sqrt{2}}\, . 
\end{align}
Therefore, the ground state $\ket{0}_+\ket{0}_-$ annihilated by $\hat{A}_\pm$ is the same as $\ket{0}_{\rm A}\ket{0}_{\rm B}$ up to a small difference of $O(c)$.
Also, up to $O(c)$ terms, $\ket{1}_+\ket{0}_-\simeq\frac{\ket{1}_{\rm A}\ket{0}_{\rm B}+\ket{0}_{\rm A}\ket{1}_{\rm B}}{\sqrt{2}}$ and $\ket{0}_+\ket{1}_-\simeq\frac{\ket{1}_{\rm A}\ket{0}_{\rm B}-\ket{0}_{\rm A}\ket{1}_{\rm B}}{\sqrt{2}}$. 

\begin{figure}[htbp]
\begin{center}
\scalebox{0.3}{
\includegraphics{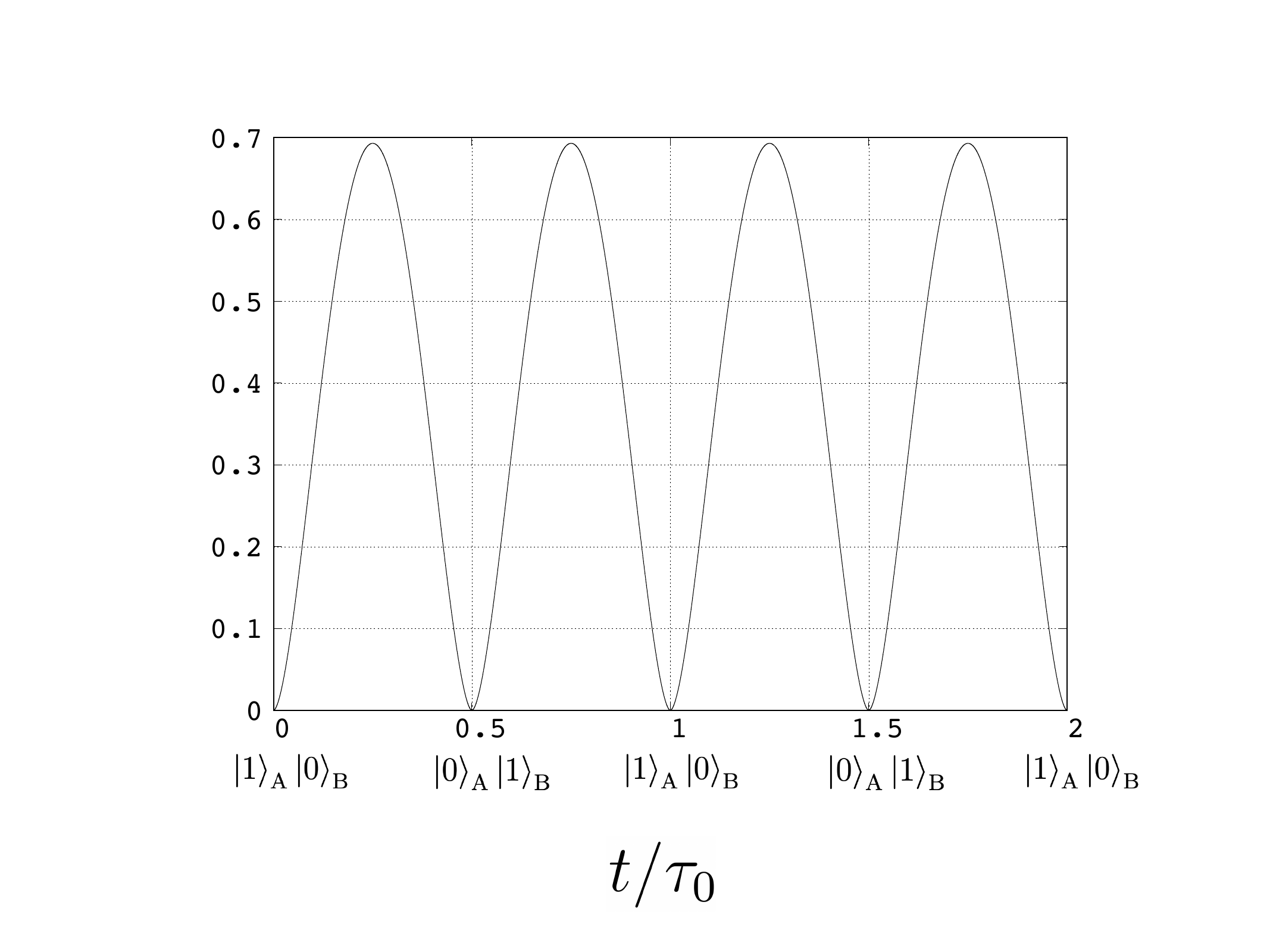}}
\end{center}
\caption{
Time evolution of entanglement entropy $-\mathrm{Tr}_A\left(\hat{\rho}_A\log\hat{\rho}_A\right)$, with the approximation \eqref{eq:EE_coupled_HO}. 
}\label{EE_coupled_HO}
\end{figure}

These expressions remind us of the low-energy wave functions of a particle in the double-well potential: $\ket{1}_+\ket{0}_-$ and $\ket{0}_+\ket{1}_-$ resembles symmetric ground state and anti-symmetric excited state. The states $\ket{1}_+\ket{0}_-$ and $\ket{0}_+\ket{1}_-$ have the energy levels $E_\pm=\omega_\pm=(1\mp c)^{1/2}$ up to the ground-state energy, and hence, the difference is $\Delta E\equiv E_--E_+\simeq c$. Because of this energy difference, there is a `tunneling' between the states localized in A or B.
Namely, time evolution of $\frac{\ket{1}_+\ket{0}_-+\ket{0}_+\ket{1}_-}{\sqrt{2}}\simeq\ket{1}_{\rm A}\ket{0}_{\rm B}$ is 
\begin{align}
&
\frac{e^{\frac{i}{2}t\Delta E}\ket{1}_+\ket{0}_-}{\sqrt{2}}
+
\frac{e^{-\frac{i}{2}t\Delta E}\ket{0}_+\ket{1}_-}{\sqrt{2}}
\nonumber\\
&\qquad\simeq
\cos\left(\pi t/\tau_0\right)\ket{1}_{\rm A}\ket{0}_{\rm B}
+
i\sin\left(\pi t/\tau_0\right)\ket{0}_{\rm A}\ket{1}_{\rm B}
\end{align} 
up to an overall phase, where $\tau_0\equiv\frac{2\pi}{\Delta E}$, and hence, it goes back and force between $\ket{1}_{\rm A}\ket{0}_{\rm B}$ and $\ket{0}_{\rm A}\ket{1}_{\rm B}$ with period $\tau_0$. 
At $\frac{t}{\tau_0}=\frac{1}{4},\frac{3}{4},\frac{5}{4},\cdots$, we have the states in the wormhole moving from A to B or from B to A:
\begin{align}
\frac{\ket{1}_{\rm A}\ket{0}_{\rm B}
\pm
i\ket{0}_{\rm A}\ket{1}_{\rm B}}{\sqrt{2}}\, . 
\end{align}
These states have entanglement analogous to \eqref{eq:A-B-entanglement}. 
The excitation can go into a wormhole due to the entanglement between regions A and B just as the example in Sec.~\ref{sec:geometry_from_entanglement}. To quantify this statement, we can employ the time evolution of entanglement entropy. At time $t$, the reduced density matrix obtained by tracing out B is
\begin{align}
\hat{\rho}_A
=
\mathrm{Tr}_{\rm B}\left(\ket{\psi(t)}\bra{\psi(t)}\right)
\simeq
\cos^2\left(\pi t/\tau_0\right)\ket{1}_{\rm A}\bra{1}_{\rm A}
+
\sin^2\left(\pi t/\tau_0\right)\ket{0}_{\rm A}\bra{0}_{\rm A}\, . 
\end{align} 
Hence, the entanglement entropy evolves as
\begin{align}
-\mathrm{Tr}_A\left(\hat{\rho}_A\log\hat{\rho}_A\right)
\simeq
-
\cos^2\left(\pi t/\tau_0\right)\cdot\log\left(\cos^2\left(\pi t/\tau_0\right)\right)
-
\sin^2\left(\pi t/\tau_0\right)\cdot\log\left(\sin^2\left(\pi t/\tau_0\right)\right)\, . 
\label{eq:EE_coupled_HO}
\end{align} 
Indeed, entanglement is large when the excitation is going through the wormhole; see Fig.~\ref{EE_coupled_HO}.
Such a wormhole is analogous to a link between two regions A and B. In this specific example, the interaction term $-c\hat{x}_{\rm A}\hat{x}_{\rm B}$ can come from $\frac{c}{2}(\hat{x}_{\rm A}-\hat{x}_{\rm B})^2$, and hence, it is analogous to the spatial derivative in the kinetic term of QFT. 

This mechanism is rather general and is applicable to various theories with discrete spectrum. For small coupling $c$, $\frac{\ket{E_n}_A\ket{E_0}_B\pm\ket{E_0}_A\ket{E_n}_B}{\sqrt{2}}$, where $\ket{E_0}$ and $\ket{E_n}$ are the ground state and the $n$-th excited state of the original system, can be close to energy eigenstates of the coupled system. Note that symmetric and antisymmetric combinations appear because of the symmetry under the exchange of system A and system B. By using these states, we can repeat the same argument presented above.\footnote{When the QFT side is gauge theory, whether the singlet constraint should be imposed or not is a nontrivial issue (see e.~g.~\cite{Maldacena:2018vsr,Jevicki:2024fnk}). This discussion is not sensitive to the singlet constraint. 
} 

Note that the interaction term is crucial for the `wormhole' to be traversable: If there is no interaction, there is no energy splitting ($\Delta E=0$), and hence, there is no `tunneling'. In the coupled SYK model or coupled SU($N$) Yang-Mills, there is a large entanglement between two sides in the ground state. The entanglement discussed above is an O($N^0$) correction to the ground state contribution coming from the small excitation. The ground state contribution assisted by `interaction' (including the spatial derivative) prepares the space that the small excitation (bulk wave packet) can go through. An important open problem is if holographic QFTs admit such nonlocal interactions at low energy that can explain the entanglement wedge reconstruction~\cite{Czech:2012bh,Wall:2012uf,Headrick:2014cta,Jafferis:2015del} (see also refs.~\cite{Almheiri:2014lwa,Dong:2016eik,Cotler:2017erl}).

Region c in Fig.~\ref{entanglement} and Fig.~\ref{entanglement_disjoint_regions} is analogous to traversable wormhole.
Region c is a bridge connecting region a and region b, and an excitation in c is created by entangling the boundary regions A and B in the same way as the above example. In the philosophy of ER$=$EPR conjecture~\cite{Maldacena:2013xja}, we could say the bulk geometry is (traversable) ER bridge connecting EPR pair (in this case, two boundary regions). 

The values of $x_{\rm A}$ and $x_{\rm B}$ are analogous to those of the scalar fields $X_I$ in 4d SYM. When the excitation is localized in region A (resp., B), the increment of the expectation value of $\hat{x}_{\rm A}^2$ (resp., $\hat{x}_{\rm B}^2$) from the ground state is 1. For the state $\frac{\ket{1}_{\rm A}\ket{0}_{\rm B}\pm i\ket{0}_{\rm A}\ket{1}_{\rm B}}{\sqrt{2}}$, because the excitation spreads to $A\cup B$, the increment from the ground state is $\frac{1}{2}$ for both $\hat{x}_{\rm A}^2$ and $\hat{x}_{\rm B}^2$. In this sense, ``radial coordinate" is smaller. This is the same as the situation in QFT: as a bulk wave packet spreads in the QFT space, scalars take smaller values. 

%%%%%%%%%%%%
%%%%%%%%%%%%
\section{Conclusion and discussion}\label{sec:conclusion}
\hspace{0.51cm}
%%%%%%%%%%%%
%%%%%%%%%%%%
In this paper, we discussed how holographic emergent geometry can be described by matrix degrees of freedom of the dual QFT or matrix model, making basic ideas provided in Refs.~\cite{Hanada:2021ipb,Hanada:2021swb,Gautam:2022akq} more concrete. Specifically, we considered the matrix model and 4d $\mathcal{N}=4$ SYM. Historically, such a direction analogous to the BFSS proposal~\cite{Banks:1996vh} was not pursued much because of a puzzling conceptual issue raised by Polchinski~\cite{Polchinski:1999br} and Susskind~\cite{Susskind:1998vk}. The key to resolving this conceptual issue was the use of wave packet in the space of matrix degrees of freedom~\cite{Hanada:2021ipb,Hanada:2021swb}, which we called matrix wave packet (see Sec.~\ref{sec:matrix-geometry-review}). 

In Sec.~\ref{sec:matrix_model}, we discussed how operators describing small excitations on semi-classical gravitational geometry can be constructed. Our construction uses only the language of the matrix model and does not assume anything about duality. The duality dictionary plays a role only when we try to interpret it in terms of gravity.  

We discussed generalization to AdS$_5$/CFT$_4$ duality in Sec.~\ref{sec:QFT}. A crucial addition to the case of the matrix model was that we must consider wave packets along the QFT space as well. Of particular importance were the global matrix wave packet and bulk wave packet, which describe the D3-brane and closed string, respectively. The bulk wave packet is a wave packet both along the QFT space and the space of matrix degrees of freedom. We found a simple interplay between the emergent space from matrix degrees of freedom and entanglement on the QFT-space that connects two seemingly different mechanisms of holographic emergent geometry, i.e., one based on matrix eigenvalues and the other based on quantum entanglement.\footnote{This combination reminds us of a tensor network approach to holography~\cite{Swingle:2009bg}.}

We did not discuss how to define the `center' of a bulk wave packet in the QFT space that would be identified with the location of the bulk excitation. It is a nontrivial issue when the support consists of multiple disconnected pieces (see e.g., Fig.~\ref{entanglement_disjoint_regions}). If the entanglement wedge reconstruction proposal is correct and if we use the location of the excitation on the gravity side to define the `center', then such a `center' may be outside the bulk wave packet. Finding a precise definition of the `center' in a purely QFT language is an important open problem.
Note also that direct and explicit computations on the QFT side are highly nontrivial. The demonstration in Sec.~\ref{sec:radial_coordinate} assumed the dual gravity description, or more precisely, a bulk reconstruction such as the HKLL~\cite{Hamilton:2006az} that uses the dual gravity description for the derivation. The demonstrations in Sec.~\ref{sec:radial-coordinate-qualitative} are essentially the dimensional analyses.

Our proposal does not immediately tell us whether the entanglement wedge reconstruction proposal is correct or wrong.\footnote{
See Refs.~\cite{Terashima:2020uqu,Sugishita:2022ldv} for criticism against the entanglement wedge reconstruction.  
} If the entanglement wedge reconstruction is valid, then the dual gravity interpretation of wave packets would depend on the boundary region under consideration. 

\begin{figure}[htbp]
\begin{center}
\scalebox{0.15}{
\includegraphics{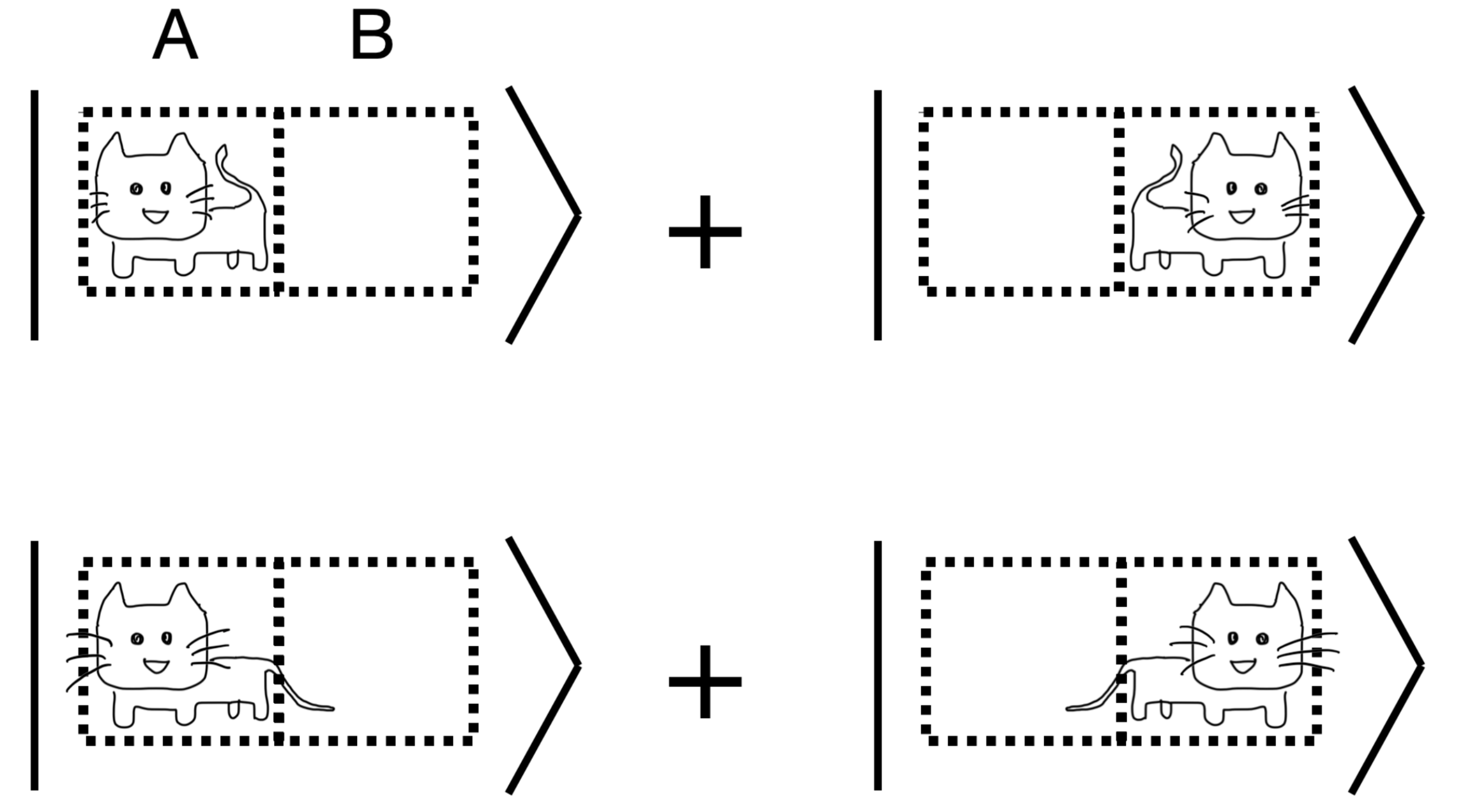}}
\end{center}
\caption{
Schematic picture of entangled states \eqref{eq:1-cat-entanglement} [top] and \eqref{eq:cat-tail-entanglement} [bottom]. 
}\label{fig:1-cat-entanglement}
\end{figure}

For the global-AdS reconstruction~\cite{Hamilton:2006az}, whole the boundary is used. Still, the argument in Sec.~\ref{sec:geometry_from_entanglement} can be applied with minor changes. An obvious change is that \eqref{eq:A-B-entanglement} cannot hold exactly; instead, we need to consider wave packets that decay outside certain regions (A$\cup$B, A, and B in this case). Accordingly, we could slightly modify \eqref{eq:1-cat-entanglement}, for example, as
\begin{align}
\frac{\ket{\rm cat\mathchar`'s\ body}_{\rm A}\otimes\ket{\rm cat\mathchar`'s\ tail}_{\rm B}+\ket{\rm cat\mathchar`'s\ tail}_{\rm A}\otimes\ket{\rm cat\mathchar`'s\ body}_{\rm B}}{\sqrt{2}}\, . 
\label{eq:cat-tail-entanglement}
\end{align}
This is still an entangled state made of one cat; see Fig.~\ref{fig:1-cat-entanglement}. 

4d $\mathcal{N}=4$ SYM has six scalars and hence not just the radial coordinate of AdS$_5$ but also S$^5$ can be described as emergent space from scalar fields. The relationship between the size of the boundary region $a$ and the `radial coordinate' $a^{-1}$ can, at least qualitatively, explain the emergence of the AdS metric. Other models could still describe the radial coordinate of AdS via the same relation although more intricate structures such as $\mathcal{M}$ of AdS$\times\mathcal{M}$ may not be captured. 

A potentially interesting application of our proposal is in the physics of small black holes in AdS$_5\times$S$^5$. 
This is an ideal setup to study black hole information problem because the evaporation of very small black holes can be studied in the isolated system~\cite{Horowitz:1999uv}. A small black hole can be described by the partially-deconfined states in the CFT~\cite{Hanada:2016pwv,Hanada:2019czd} (see also Fig.~\ref{matrix-1-excitation-PD}). The emission of graviton can be visualized as the emission of bulk wave packets. The size of the deconfined sector decreases as the black hole shrinks. This picture provides us with an intuitive understanding of important properties such as negative specific heat~\cite{Berkowitz:2016znt,Hanada:2016pwv} and Page curve~\cite{Gautam:2022akq}. An important open problem is how the event horizon can be understood in terms of matrix model or QFT.\footnote{An intuitive picture could be as follows. 
For a small block to be emitted from the deconfined sector (black hole), the off-diagonal entries of $Y$ and $Q$ connecting the deconfined sector and small block must vanish. Such a process is entropically suppressed; as the deconfined sector becomes larger, the emission is more suppressed. This could explain the formation of the event horizon.
When a small block falls into the deconfined sector, off-diagonal entries are excited. The dual interpretation could be the condensation of open strings.
} 

Conceptually the simplest test and potentially the most powerful application are in quantum simulations. If we can simulate the matrix model or SYM on a quantum computer, we can simply monitor the time evolution of the scalar field and identify which degrees of freedom describe which object. 
%%%%%%%%%%%%%%%%%%%%%%
%%%%%%%%%%%%%%%%%%%%%%
\begin{center}
\Large{\textbf{Acknowledgement}}
\end{center}
%%%%%%%%%%%%%%%%%%%%%%
%%%%%%%%%%%%%%%%%%%%%%
The authors would like to thank Tarek Anous, Masazumi Honda, Dan Kabat, Henry Lin, Hong Liu, So Matsuura, Cheng Peng, Hidehiko Shimada, Evgeny Sobko, and Brian Swingle for discussions and comments. V.~G. thanks STFC for the Doctoral Training Programme funding (ST/W507854-2021 Maths DTP). M.~H. thanks the STFC for their support through the consolidated grant ST/Z001072/1. A.~J. was supported by the U.S. Department of Energy under contract DE-SC0010010. M.~H. and A.~J. thank the participants of the APCTP Focus Program ``Entanglement, Large $N$, and Black Hole 2024" for questions, comments, and discussions during and after M.~H.'s presentation.

\appendix
%%%%%%%%%%%%
%%%%%%%%%%%%
\section{Polchinski's puzzle}\label{sec:Polchinski-puzzle}
\hspace{0.51cm}
%%%%%%%%%%%%
%%%%%%%%%%%%
Let us explain a seemingly puzzling conceptual issue on the emergent geometry from matrix degrees of freedom~\cite{Polchinski:1999br,Susskind:1998vk} by using a matrix model with the Hamiltonian \eqref{eq:toy-Hamiltonian} as an example. 
The following argument uses only generic properties of large-$N$ gauge theory. Therefore, essentially the same puzzle existed for AdS$_5$/CFT$_4$ as well; see e.g.~\cite{Heemskerk:2012mn}. Note also that, though we will focus on the 't Hooft large-$N$ limit, essentially the same results are obtained quite generally including the parameter region of the BFSS matrix model that describes M-theory.

The simple relations that created huge, decades-long confusion are
\begin{align}
\bra{\Phi}\mathrm{Tr}\hat{X}_{I}^2\ket{\Phi}\sim N^2
\end{align}
and
\begin{align}
\bra{\Phi}\mathrm{Tr}[\hat{X}_{I},\hat{X}_J]^2\ket{\Phi}\sim N^3\, . 
\end{align}
These are consequences of 't Hooft scaling and hence hold quite generally for any theory and any state $\ket{\Phi}$ whose energy is of order $N^2$ or smaller. In particular, these relations hold for the ground state as well. 

Let us imagine we performed a measurement in the coordinate basis and obtained a generic $X_{I,ij}$. In retrospect, we should have known that such a $X_{I,ij}$ does not necessarily carry the information of emergent geometry --- we must take a very special point, i.e., the center of wave packet --- but suppose that such matrices $X_{I,ij}$ could describe geometry and see what kinds of `physics' would follow. Note that such an assumption could look natural, given an apparent analogy to the master field. Then, such typical matrices satisfy
\begin{align}
\mathrm{Tr}X_{I}^2\sim\bra{\Phi}\mathrm{Tr}\hat{X}_{I}^2\ket{\Phi}\sim N^2\, , 
\end{align}
and hence, eigenvalues of each $X_I$ are typically of order $\sqrt{N}$, which is as large as the radius of the bulk region where the weakly-curved gravity picture is valid. This is the case even for the ground state. It looks as though the D-branes are fluctuating a lot, and the wave function is not localized and the resolution for the `location' is so bad that it cannot capture the geometry on the gravity side. Furthermore, if we diagonalize $X_1$, then 
\begin{align}
\mathrm{Tr}[X_{1},X_J]^2\sim\bra{\Phi}\mathrm{Tr}[\hat{X}_{1},\hat{X}_J]^2\ket{\Phi}\sim N^3
\end{align}
for $J\neq 1$, which means 
\begin{align}
\sum_{i\neq j}(X_{1,ii}-X_{1,jj})^2|X_{J,ij}|^2
\sim
N^3\, . 
\end{align}
Because $(X_{1,ii}-X_{1,jj})^2$ is typically of order $N$, each $|X_{J,ij}|$ is of order 1. 
Given that $\mathrm{Tr}X_J^2\sim N^2$, a significant fraction of $\mathrm{Tr}X_J^2$ comes from the off-diagonal entries of $X_J$. 
In other words, when $X_1$ is diagonalized, other $X_J$'s are not close to diagonal. For this reason, it looks as though the basic assumption of the geometric picture --- $X_I$'s can be taken close to simultaneously diagonal --- does not hold. 

Because this pathological feature appears even for the ground state, it was clear that the quantum fluctuation must be removed in order to obtain the right geometric information~\cite{Polchinski:1999br}. Ref.~\cite{Hanada:2021ipb} pointed out that the geometric information is obtained from the center of the wave packet..  
%%%%%%%%%%%%
%%%%%%%%%%%%
\section{Computation of three-point function in Sec.~\ref{sec:radial_coordinate}}\label{sec:3pt-function}
\hspace{0.51cm}
%%%%%%%%%%%%
%%%%%%%%%%%%
In this appendix, we use $z=r^{-1}$ frequently. 
As $\hat{O}$, we take chiral primaries that consist of $k_O$ scalar fields. 
We use $\hat{O}_{r,\vec{x}}$ and $\hat{O}_{z,\vec{x}}$ to mean the same operator. 
To regularize the short-distance divergence in the bulk, we can use Gaussian wave packet in the bulk:\footnote{
Note also that the examples discussed in Sec.~\ref{sec:radial-coordinate-qualitative} correspond to such a wave packet on the gravity side with a finite value of $\sigma_0$. 
}
\begin{align}
\hat{O}^{\rm (smear)}_{z,\vec{x}}
=
\frac{1}{4\pi^2\sigma^4(z)}
\int d^3\vec{x}'dz' e^{-\frac{(z-z')^2+(\vec{x}-\vec{x}')^2}{2\sigma^2(z)}}
\hat{O}_{z',\vec{x}'}\, . 
\end{align}
We take $\sigma$ to vary with the radial coordinate $z$ as
\begin{align}
\sigma(z)=\sigma_0\times z\, , 
\end{align}
where $\sigma_0$ is a constant, so that the spatial volume of the wave packet is fixed. 
We take the variance small enough such that $\sigma_0\ll 1$. The two-point function of $\hat{O}^{\rm (smear)}_{z,\vec{x}}$ at the same point is 
\begin{align}
    \frac{1}{(4\pi^2\sigma^4)^2}
    \int d^3\vec{x}_1dz_1 e^{-\frac{(z-z_1)^2+(\vec{x}-\vec{x}_1)^2}{2\sigma^2}}
    \int d^3\vec{x}_2dz_2 e^{-\frac{(z-z_2)^2+(\vec{x}-\vec{x}_2)^2}{2\sigma^2}}
    \langle\textrm{g.s.}| \hat{O}^\dagger_{z_1,\vec{x}_1} \hat{O}_{z_2,\vec{x}_2} |\textrm{g.s.}\rangle\, . 
    \label{eq:bulk-to-bulk-2pt-smeared}
\end{align}
The \textit{bulk}-to-\textit{bulk} propagator in AdS$_5$ for a massive scalar field $\hat{O}$ of mass squared $m^2 = k_O (k_O - 4)$ is given by, up to an overall factor, \footnote{Note that the one-point function vanishes for chiral primaries.}~\cite{burgess1985propagators,Freedman:1998tz}
\begin{align}
    \langle\textrm{g.s.}| \hat{O}^\dagger_{r_1,\vec{x}_1} \hat{O}_{r_2,\vec{x}_2} |\textrm{g.s.}\rangle = 
    \xi^{k_O} F\left(\frac{k_O}{2},\frac{k_O+1}{2};k_O-1;\xi^2\right)\, ,  
    \label{eq:bulk-to-bulk-2pt}
\end{align}
where $F$ is the hypergeometric function and 
\begin{align}
    \xi = \frac{2z_1z_2}{z_1^2 + z_2^2 + (\vec{x}_1-\vec{x}_2)^2}\, . 
\end{align}
To regularize the denominator of \eqref{eq:3-pt-fnc}, we replace $\hat{O}$ with $\hat{O}^{\rm (smear)}$. We take $\sigma_0$ parametrically small such that only $z_1,z_2\simeq z$ and $\vec{x}_1,\vec{x}_2\simeq\vec{x}$ contribute. Then, $\xi$ can be approximated as
\begin{align}
\xi
\simeq
\left(1+\frac{(z_1-z_2)^2+(\vec{x}_1-\vec{x}_2)^2}{2z^2}\right)^{-1}\, . 
\end{align}
Then, $F\left(\frac{k_O}{2},\frac{k_O+1}{2};k_O-1;\xi^2\right)$ diverges as $(1-\xi^2)^{-3/2}\sim\left(\frac{z^2}{(z_1-z_2)^2+(\vec{x}_1-\vec{x}_2)^2}\right)^{3/2}$.
Therefore, \eqref{eq:bulk-to-bulk-2pt-smeared} becomes
\begin{align}
&
    \frac{1}{(4\pi^2\sigma^4)^2}
    \int d^3\vec{x}_1dr_1 e^{-\frac{(z-z_1)^2+(\vec{x}-\vec{x}_1)^2}{2\sigma^2}}
    \int d^3\vec{x}_2dr_2 e^{-\frac{(z-z_2)^2+(\vec{x}-\vec{x}_2)^2}{2\sigma^2}}
    \left(\frac{z^2}{(z_1-z_2)^2+(\vec{x}_1-\vec{x}_2)^2}\right)^{3/2}
    \nonumber\\
    &=
    \frac{1}{16\pi^2\sigma^4}
    \int d^3\vec{\delta}d\epsilon e^{-\frac{\epsilon^2+\vec{\delta}^2}{4\sigma^2}}
    \left(\frac{z^2}{\epsilon^2+\vec{\delta}^2}\right)^{3/2}\, ,     
\end{align}
where $\epsilon=z_1-z_2$ and $\vec{\delta}=\vec{x}_1-\vec{x}_2$, and hence, by using $\rho=\sqrt{\epsilon^2+\vec{\delta}^2}$,
\begin{align}
    \frac{1}{16\pi^2\sigma^4}
    \int_0^\infty 2\pi^2\rho^3d\rho e^{-\frac{\rho^2}{4\sigma^2}}
    \frac{z^3}{\rho^3}
    =
   \frac{\sqrt{\pi}z^3}{4\sigma^3} 
       =
   \frac{\sqrt{\pi}}{4\sigma_0^3}\, . 
   \label{eq:normalization-2-pt}
\end{align}

We use the same regularization for the three-point function. Then, the numerator of \eqref{eq:3-pt-fnc} becomes the product of three-point vertex, two bulk-to-bulk propagators \eqref{eq:bulk-to-bulk-2pt}, and 
bulk-to-boundary propagator, integrated over the location of the three-point vertex which we denote by $\tau_3,r_3,\vec{x}_3$. 
Because the dominant contribution comes from $\tau_3\sim 0$, $r_3\sim r$ and $\vec{x}_3\sim\vec{x}$, the bulk-to-boundary propagator for the rank-$k$ chiral primary operator is, up to a normalization factor~\cite{Freedman:1998tz} and negligible corrections, 
\begin{align}
    K(r,\vec{x};\vec{x}') = \left( \frac{r^{-1}}{r^{-2} + (\vec{x}-\vec{x}')^2}\right)^k\, . 
    \label{eq:bulk-to-boundary}
\end{align}
This sets the $r$-dependence of the three-point function. Note that this scaling is valid for generic local operators in the bulk. 
The divergence comes from the product of two bulk-to-bulk propagators integrated over the location of three-point vertex,
\begin{align}
&
    \frac{1}{(4\pi^2\sigma^4)^2}
    \int d^3\vec{x}_1dz_1 e^{-\frac{(z-z_1)^2+(\vec{x}-\vec{x}_1)^2}{2\sigma^2}}
    \int d^3\vec{x}_2dz_2 e^{-\frac{(z-z_2)^2+(\vec{x}-\vec{x}_2)^2}{2\sigma^2}}
        \nonumber\\
    &\qquad
        \int\frac{d^3\vec{x}_3dz_3d\tau_3}{z^5}
    \left(\frac{z^2}{z_{13}^2+\vec{x}_{13}^2+\tau_3^2}\right)^{3/2}
    \left(\frac{z^2}{z_{23}^2+\vec{x}_{23}^2+\tau_3^2}\right)^{3/2}
    \label{eq:2-bulk-to-bulk-propagators}
\end{align}
where $z_{13}=z_1-z_3$, etc. To evaluate the second line, let us introduce $\tau_1$ and $\tau_2$, which are sent to zero later, and write the integral by using a 5-component vector $\boldsymbol{x}=(\tau,z,\vec{x})$ as
\begin{align}
      z\int d^3\vec{x}_3dz_3d\tau_3
    \left(\frac{1}{z_{13}^2+\vec{x}_{13}^2+\tau_{13}^2}\right)^{3/2}
        \left(\frac{1}{z_{23}^2+\vec{x}_{23}^2+\tau_{23}^2}\right)^{3/2}
    =
    z\int \frac{d^5\boldsymbol{x}_3}{|\boldsymbol{x}_{13}|^3|\boldsymbol{x}_{23}|^3}\, . 
\end{align}
This integral depends only on the distance $|\boldsymbol{x}_{12}|$ and, by the dimensional counting, we obtain $\frac{z}{|\boldsymbol{x}_{12}|}$ up to a multiplicative factor. Plugging it back to \eqref{eq:2-bulk-to-bulk-propagators}, we obtain $\frac{z}{\sigma}=\frac{1}{\sigma_0}$. Combining this factor, \eqref{eq:normalization-2-pt}, and \eqref{eq:bulk-to-boundary}, we obtain
\begin{align}
\frac{
\bra{\rm g.s.}\hat{O}^{\rm (smeared)\dagger}_{r,\vec{x}}
\mathrm{Tr}(X_{\{I_1}\cdots X_{I_k\}})(\vec{x}')
\hat{O}^{\rm (smeared)}_{r,\vec{x}}\ket{\rm g.s.}_{\rm conn}
}{
\bra{\rm g.s.}\hat{O}^{\rm (smeared)\dagger}_{r,\vec{x}}\hat{O}^{\rm (smeared)}_{r,\vec{x}}\ket{\rm g.s.}
}
\sim
    \sigma_0^2\times\left( \frac{r^{-1}}{r^{-2} + (\vec{x}-\vec{x}')^2}\right)^k\, . 
\end{align}
This gives us the expected scaling explained in Sec.~\ref{sec:radial_coordinate}. That the result depends on the size of the wave packet through the overall factor $\sigma_0^2$ may look strange at first sight, but that it vanishes as the wave packet $\hat{O}^{\rm (smeared)}\ket{\rm g.s.}$ shrinks is actually natural. The three-point function can be interpreted as the overlap of two states $\mathrm{Tr}(\hat{X}_{\{I_1}\cdots\hat{X}_{I_k\}})(\vec{x}')\hat{O}^{\rm (smeared)}\ket{\rm g.s.}$ and $\hat{O}^{\rm (smeared)}\ket{\rm g.s.}$. While the latter is localized in the bulk by assumption, we expect that the former is not because there is no reason to be so. (Note that $\hat{O}^{\rm (smeared)}\ket{\rm g.s.}$ is not necessarily an eigenstate of $\mathrm{Tr}(X_{\{I_1}\cdots X_{I_k\}})(\vec{x}')$.) Hence the overlap should become smaller as the wave packet shrinks. 

The normalization of a three-point vertex can depend on the normalization of two-point functions. When the two-point function does not have a $k$-dependent overall factor, the three-point vertex for rank-$k_1$, $k_2$ and $k_3$ chiral primary is proportional to $\sqrt{k_1k_2k_3}$~\cite{Lee:1998bxa}. For $k_1=k_2=k_O$, this is proportional to $k_O$. This factor $k_O$ appears in the three-point function normalized as \eqref{eq:3-pt-fnc}.

%%%%%%%%%%%%%%%%%%%%%%%%%
%\bibliographystyle{unsrt}
\bibliographystyle{utphys}
\bibliography{wave_packet}

\end{document}